\documentclass[10pt]{article}

\usepackage{amsmath}
\usepackage{amssymb}

\usepackage{graphicx}

\usepackage{cite}

\usepackage{color} 

\usepackage[labelfont=bf,labelsep=period,justification=raggedright]{caption}
\usepackage{subcaption}
\makeatletter
\renewcommand{\@biblabel}[1]{\quad#1.}
\makeatother

\date{}

\usepackage{hyperref}

\graphicspath{{./figures/}}

\begin{document}

\begin{flushleft}
{\Large
\textbf{Measuring Large-Scale Social Networks with High Resolution}
}
\\
Arkadiusz Stopczynski$^{1,\ast}$, 
Vedran Sekara$^{1}$, 
Piotr Sapiezynski$^{1}$, 
Andrea Cuttone$^{1}$,
Mette My Madsen$^{3}$ 
Jakob Eg Larsen$^{1}$, 
Sune Lehmann$^{1,2}$
\\
\bf{1} Technical University of Denmark\\
\bf{2} The Niels Bohr Institute\\
\bf{3} University of Copenhagen
\\
$\ast$ E-mail: arks@dtu.dk
\end{flushleft}

\tableofcontents
\newpage
\section{Abstract}

This paper describes the deployment of a large-scale study designed to measure human interactions across a variety of communication channels, with high temporal resolution and spanning multiple years---the Copenhagen Networks Study. 
Specifically, we collect data on face-to-face interactions, telecommunication, social networks, location, and background information (personality, demographic, health, politics) for a densely connected population of 1\,000 individuals, using state-of-art smartphones as social sensors.
Here we provide an overview of the related work and describe the motivation and research agenda driving the study. 
Additionally the paper details the data-types measured, and the technical infrastructure in terms of both backend and phone software, as well as an outline of the deployment procedures.
We document the participant privacy procedures and their underlying principles.
The paper is concluded with early results from data analysis, illustrating the importance of multi-channel high-resolution approach to data collection.


\section{Introduction}

Driven by ubiquitous availability of data and inexpensive data storage capabilities, the concept of big data has permeated the public discourse and led to surprising insights across the sciences and humanities~\cite{ginsberg2008detecting, aral2012identifying}.
While collecting data may be relatively easy, it is a challenge to combine datasets from multiple sources.
In part this is due to mundane practical issues, such as matching up noisy and incomplete data, and in part due to complex legal and moral issues connected to data ownership and privacy, since many datasets contain sensitive data regarding individuals~\cite{IMM2013-06632}.
As a consequence, most large datasets are currently locked in `silos', owned by governments or private companies, and in this sense the big data we use today are `shallow'---only a single or very few channels are typically examined.

Such shallow data limit the results we can hope to generate from analyzing these large datasets.
We argue below (in~\nameref{sec:motivation} section) that in terms of understanding of human social networks, such shallow big data sets are not sufficient to push the boundaries in certain areas.
The reason is that human social interactions take place across various communication channels; we seamlessly and routinely connect to the same individuals using face-to-face communication, phone calls, text messages, social networks (such as Facebook and Twitter), emails, and many other platforms.
Our hypothesis is that in order to understand social networks, we must study communication across these many channels which are currently siloed.
Existing big data approaches have typically concentrated on large populations ($\mathcal{O}(10^5) - \mathcal{O}(10^8)$), but with a relatively low number of bits per user, for example in call detail records (CDR) studies~\cite{onnela_structure_2007} or Twitter analysis~\cite{cha2010measuring}.
Here, we are interested in capturing deeper data, looking at multiple channels from sizable populations. 
Using big data collection and analysis techniques, which can scale in number of users, we show how to start deep, i.e. with detailed information about every single study participant, and then scale up to very large populations.

We are not only interested in collecting deep data for a large, highly connected population, but also aim to create a dataset collected in an interactive way, where we can change the collection process.
This enables us to rapidly adapt and change our collection methods if current data, for example, have insufficient temporal resolution for a certain question we would like to answer. 
We have designed our data collection setup in such a way we are able to deploy experiments, since we know that causal inference is notoriously complicated in network settings~\cite{shalizi2011homophily}, and perform continuous quality control of the data collected.
The mindset of real-time data access can be extended beyond pure research, where the goal is to monitor data quality and perform interventions. 
Using the methods described here, we can potentially use big data in real-time to observe and react to the processes taking place across entire societies.
In order to achieve this goal, researchers must approach the data in the same way large Internet services do---as a resource that can be manipulated and made available in real-time, since it inevitably loses value over time.

In order to realize the interactive data collection, we need to build long-lasting testbeds to rapidly deploy experiments, while still retaining access to all the data collected hitherto.
Human beings are not static; our behavior, our networks, our thinking changes over time~\cite{barabasi2002evolution, kossinets2006empirical}.
To be able to analyze and understand changes over long time scales, we need longitudinal data, available not just to a single group of researchers, but to changing teams of researchers, working with an evolving set of ideas, hypotheses, and perspectives. 
Ultimately, we aim to be able to access the data containing the entire life-experience of people and look at their lives as dynamic processes.
Eventually, we aim to even go beyond the lifespan of individuals and analyze the data of the entire generations. 
We are not there yet, but mankind is moving in this direction: all tweets are archived in the Library of Congress (\url{https://blog.twitter.com/2010/tweet-preservation}), a person born today in a developed country has a good chance of keeping every single picture they ever take, the next generation will have a good chance of living with highly detailed lifelog, such as every single electronic message they have ever exchanged with their friends.
The status quo is we need to actively opt out for our experiences not to be auto-shared: major cloud storage providers offer auto-upload feature for pictures taken with a smartphone, every song we listen to on Spotify is remembered and used to build our profile---unless we explicitly turn on private mode. 

In this paper we describe a large-scale study observing student lives through multiple channels---the Copenhagen Network Study.
With an iterative approach to deployments, it is an example of an interdisciplinary approach.
We collect data from multiple sources, including questionnaires, online social networks, and smartphones handed out to students.
Data from all those channels create multi-layered view of the individuals, their networks, and their environments; those views can be then examined separately and jointly by researchers of different expertise.
We are building the Copenhagen Network Study as a framework for long-lived extensible studies. The 2012 and 2013 deployments described here are called \emph{SensibleDTU} and are based at the Technical University of Denmark; they have been designed as part of a close collaboration with researchers from the social sciences, natural sciences, medicine (public health), and humanities, as part of the \emph{Social Fabric} project (see Acknowledgements for details).
Currently in the second iteration, we have deployed phones for about 1\,000 participants, growing a dataset of unprecedented size and resolution.
In addition to the core task of collecting deep behavioral data, we also experiment with creating rich services for our participants and improving privacy practices.

Human lives, especially when seen in the perspective of months and years, take place in multiple dimensions. 
Capturing only a single channel, even for the entire persons' life, limits the expertise that can be applied to understand a human being.
True interdisciplinary studies require deep data. 
Anthropologists, economists, philosophers, physicists, psychologists, public health researchers, sociologists, and computational social science researchers are all interested in distinct questions, and traditionally use very different methods.
We believe that it is when those groups start working together, qualitatively better findings can be made. 

Here we give a brief overview of the related work, in the domains of data collection and analysis, extend the description of the motivation driving the project, and outline the experimental plan and data collection methodology.
We report on the implemented privacy and informed consent practices, emphasizing how we went beyond the usual practice in such studies, and created some cutting edge solutions in the domain.
We also report a few initial results from the project, primarily in the form of the overview of collected data, and outline future directions.
We hope the work presented here will serve as a guideline for deploying similar massive sensor-driven human data collection studies.
With the overview of the collected data, we extend an invitation to researches of all fields to contact the authors for the purpose of defining novel projects around the Copenhagen Networks Study testbed.

\section{Related Work}

Lazer et al.~introduced computational social science (CSS) as a new field of research that studies individuals and groups in order to understand populations, organizations, and societies using big data---phone call records, GPS traces, credit card transactions, webpage visits, emails, data from social networks~\cite{lazer2009life}.
CSS focuses on questions that can now be studied using data-driven computational analyses of datasets such as the ones mentioned above, and which could only previously be addressed self-reported data or direct observations: dynamics in work groups, face-to-face interactions, human mobility, or information spread.
The hope is that such a data-driven approach will bring new types of insight, not available with classical methods.
The challenges that emerge in this set of new approaches, include wrangling big data, applying network analysis to dynamic networks, ensuring privacy of personal information, and enabling interdisciplinary work between computer science and social science, to name just a few.

In this section we describe related work in terms of the central methods of data collection, provide a brief overview of results obtained from the analysis of CSS data, and mention some principles regarding privacy and data treatment.

\subsection{Data collection}
Many of the CSS studies have been performed on call detail records (CDRs)---records of user phone calls and messages collected by mobile phone operators.
Although CDRs can be a proxy for mobility and social interaction~\cite{wesolowski2013impact}, much of the social interaction happens face-to-face, and may be difficult to capture with CDRs or other channels such as social networks (Twitter, Facebook, etc.)~\cite{madrigal2013dark}.
To gain a fuller view of participants' behavior, some CSS studies have developed an approach of employing Radio Frequency Identification (RFID) devices~\cite{cattuto2010dynamics}, sociometetric badges~\cite{wu2008mining, polastre2005telos}, as well as smartphones for the data collection~\cite{raento2009smartphones, chronis2009socialcircuits,pentland2008honest,olguin2011mobile}.
Smartphones are unobtrusive, relatively cheap, feature a plethora of embedded sensors, and tend to travel nearly everywhere with their users.
They allow for automatic collection of sensor data including GPS, Wifi, Bluetooth, calls, SMS, battery, and application usage~\cite{miller2012smartphone}.
Collecting data with smartphones presents, however, several limitations: sensing is mainly limited to pre-installed sensors, which may not be of highest quality, and off-the-shelf software and hardware may not be sufficiently robust for longitudinal studies. 

A large number of solutions for sensor-driven human data collection have been developed, ranging from dedicated software to complete platforms, notably ContextPhone~\cite{raento2005contextphone}, SocioXensor~\cite{mulder2005socioxensor},~MyExperience \cite{froehlich2007myexperience}, Anonysense~\cite{cornelius2008anonysense}, CenceMe~\cite{miluzzo2008sensing}, Cityware~\cite{kostakos2008cityware}, Darwin phones~\cite{miluzzo2010darwin}, Vita~\cite{hu2013vita}, and ContextToolbox~\cite{larsen2009mobile}.

Running longitudinal rich behavioral data collection from large populations presents multiple logistical challenges and only few studies have attempted it so far.
In the Reality Mining study, data from 100 mobile phones were collected for nine months~\cite{eagle2006reality}.
In the Social fMRI study, 130 participants carried smartphones running the Funf mobile software~\cite{funf_website} for 15 months~\cite{aharony2011social}.
Data was also collected from Facebook, credit card transactions, and surveys were pushed to the participants' phones.
The Lausanne Data Collection Campaign~\cite{kiukkonen2010towards,laurila2012mobile} featured 170 volunteers in the Lausanne area of Switzerland, between October 2009 and March 2011.
In the SensibleOrganization study~\cite{olguin2009sensible} researchers used RFID tags for a period of one month to collect face-to-face interactions of 22 employees working in a real organization.
Preliminary results with 20 participants from a large university campus have been so far reported from the OtaSizzle study~\cite{karikoski2011measuring}, and in the Locaccino study~\cite{cranshaw2010bridging}, location within a metropolitan region was recorded for 489 users for varying periods, ranging from seven days to several months.

\subsection{Data analysis}
Below we provide selected examples of results obtained from analysis of CSS datasets in various domains.

\subsubsection{Human Mobility}

Gonzales et al.~analyzed six months of CDRs of 100\,000 users, finding that human mobility is quite predictable, with high spatial and temporal regularity, and few highly frequented locations~\cite{gonzalez2008understanding}. 
This result was extended by Song et al., where three months of CDRs from 50\,000 individuals were analyzed, finding a 93\% upper bound of predictability of human mobility, a number which applies to most users regardless of different travel patterns and demographics~\cite{song2010limits}.
Sevtsuk et al.~focused instead on the aggregate usage of 398 cell towers, describing the hourly, daily, and weekly patterns and their relation to demographics and city structure~\cite{demographics}. 
Bagrow et al.~analyzed 34 weeks of CDRs for 90\,000 users, identifying habitats---groups of related places---and finding that most individuals in their dataset had between 5 and 20 habitats~\cite{bagrow2012mesoscopic}.
De Domenico et al.~showed in~\cite{de2012interdependence} how location prediction can be performed using multivariate non-linear time series prediction, and how accuracy can be improved considering the geo-spatial movement of other users with correlated mobility patterns.

\subsubsection{Social Interactions}

Face-to-face interactions can be used to model social ties over time and organizational rhythms in response to events~\cite{eagle2006reality, eagle2009inferring, eagle2009eigenbehaviors}.
Comparing those interactions with Facebook networks, Cranshaw et al.~found that meetings in locations of high entropy (featuring diverse set of visitors) are less indicative than meetings in locations visited by a small set of users~\cite{cranshaw2010bridging}. 
Clauset et al.~found that a natural time scale of face-to-face social networks is $4$ hours~\cite{clauset2012persistence}. 
 
Onnela et al.~analyzed CDRs from 3.9 millions users~\cite{onnela2007analysis}, finding evidence supporting the weak ties hypothesis~\cite{granovetter1973strength}.
Lambiotte et al.~analyzed CDRs from 2 millions user, finding that the probability of the existence of the links decreases as $d^{-2}$, where $d$ is the distance between users~\cite{lambiotte2008geographical}. 
In another study with CDRs from 3.4 million users, the probability was found to decrease as $d^{-1.5}$~\cite{onnela2011geographic}.
Hidalgo et al.~found---analyzing CDRs for 2 millions users---that persistent links tend to be reciprocal and associated with low degree nodes~\cite{hidalgo2008dynamics}.

Miritello et al.~analyzed CDRs for 20 millions people, and observed that individuals have a finite limit of number of active ties, and two different strategies for social communication~\cite{miritello2013limited, miritello2013time}.
Sun et al.~analyzed 20 million bus trips from about 55\% of the Singapore population and found distinct temporal patterns of regular encounters between strangers, resulting in a co-presence network across the entire metropolitan area~\cite{sun2013understanding}.

\subsubsection{Health and Public Safety}

Using CDRs from the period of the 2008 earthquake in Rwanda, Kapoor et al.~created a model for detection of the earthquake, the estimation of the epicenter, and determination of regions requiring relief efforts~\cite{kapoor2010people}.
Aharony et al.~performed and evaluated a fitness activity intervention with different reward schemes, based on face-to-face interactions~\cite{aharony2011social} while Madan et al.~studied how different illnesses (common cold, depression, anxiety) manifest in common mobile-sensed features (Wifi, location, Bluetooth) and the effect of social exposure on obesity~\cite{madan2012sensing}. 
Salath\'{e} et al.~showed that disease models simulated on top of proximity data obtained from a high school are in good agreement with absentees from an influenza season \cite{salathe2010high}, and emphasize that contact data is required to design effective immunization strategies. 

\subsubsection{Influence and Information Spread}

Chronis et al.~\cite{chronis2009socialcircuits} and Madan et al.~\cite{madan2011pervasive} investigated how face-to-face interactions affect political opinions.
Wang et al.~reported on the spread of viruses in mobile networks; Bluetooth viruses can have a very slow growth but can spread over time to a large portion of the network, while MMS viruses can have an explosive growth but their spread is limited to subnetworks~\cite{wang2009understanding}.
Aharony et al.~analyzed the usage of mobile apps in relation with face-to-face interactions, finding that more face-to-face interaction increases the number of common applications~\cite{aharony2011social}.
Using RFID for sensing face-to-face interactions, Isella et al.~estimated the most probable vehicles for infection propagation~\cite{isella2011close}.
With a a similar technique, but applied to 232 children and 10 teachers in a primary school, Stehle et al. described a strong age homophily in children interactions~\cite{stehle2011high}.

Bagrow et al.~showed how CDRs communications in relation to entertainment events (e.g.~concerts, sporting events) and emergencies (e.g.~fires, storms, earthquakes) have two well-distinguishable patterns in human movement~\cite{bagrow2011collective}.
Karsai et al.~analyzed CDR from six millions users to find that strong ties tend to constrain the information spread within localized groups of individuals~\cite{karsai2013emergence}.

Studies of Christakis and Fowler on obesity and smoking spread in networks~\cite{christakis2007spread, christakis2008collective} prompted a lively debate on how homophily and influence are confounded, with Lyons criticizing used statistical methods~\cite{lyons2011spread}.
Stelich et al.~discussed how friendship formation in a dynamic network based on homophily can be mistaken for influence~\cite{steglich2010dynamic} and Shalizi and Thomas showed examples of how homophily and influence can be confounded~\cite{shalizi2011homophily}.
Finally, Aral et al.~provided a generalized statistical framework for distinguishing peer-to-peer influence from homophily in dynamic networks~\cite{aral2009distinguishing}.

\subsubsection{Socioeconomics and Organizational Behavior}

Face-to-face contacts and email communication for employees in a real work environment can be used to predict job satisfaction and group work quality~\cite{olguin2009sensible}. 
Having more diverse social connections is correlated with economics opportunities, as found in the study containing CDRs of over 65 millions users~\cite{eagle2010network}.
A similar result was reported in a study of economic status and physical proximity, where a direct correlation between more social interaction diversity and better financial status was found~\cite{aharony2011social}.  
Or, as shown in a study of Belgian users, language regions in a country can be identified based solely on CDRs~\cite{blondel2010regions}.

\subsection{Privacy}

Data collected about human participants is sensitive and ensuring privacy of the users is a fundamental requirement---even when those users may have limited understanding of the implications of data sharing~\cite{mahato2008implicit, klasnja2009exploring}.
A significant literature exists regarding the possible attacks that can be performed on personal data, such as unauthorized analysis of personal data~\cite{altshuler2011stealing} to decode daily routines~\cite{shokri2011quantifying} or friendships~\cite{eagle2009inferring} of the participants. 
In \emph{side channel information} attacks data from public datasets (e.g.~online social networks) are used to re-identify users~\cite{lane2012feasibility, srivatsa2012deanonymizing, mislove2010you}. 
Even connecting different records of one user within the same system can compromise privacy~\cite{lane2012feasibility}. 
Specific attacks are also possible in network data, as nodes can be identified based on the network structure and attributes of the neighbors~\cite{zhou2008preserving, cheng2010k}.

Various de-identification techniques can be applied to the data.
\emph{Personally Identifiable Information} (PII) is any information that can be used to identify an individual, such as name, address, social security number, date and place of birth, employment, education, or financial status. 
In order to avoid re-identification and consequent malicious usage of data, PIIs can be completely removed, hidden by aggregation, or transformed to be less identifiable, resulting in a trade-off between privacy and utility~\cite{li2009tradeoff}.
Substituting PII with the correspondent one-way hash allows to remove plaintext information and break the link to other datasets. 
This method, however, does not guarantee protection from re-identification~\cite{narayanan2008robust, sweeney2000simple, barbaro2006face, de2013unique}.
$K-$anonymity is a technique of ensuring that it is not possible to distinguish any user from at least $k-1$ other in the dataset~\cite{sweeney2002k}; studies have shown that this method may be often too weak~\cite{shokri2011quantifying}.
$L-$diversity~\cite{machanavajjhala2007diversity} and $t-$closeness~\cite{li2007t} have been proposed as extensions of $k-$anonymity with stronger guarantees.

Another approach to introducing privacy is based on perturbing the data by introducing noise, with the goal of producing privacy-preserving statistics~\cite{dinur2003revealing, dwork2004privacy, blum2005practical, dwork2006our, chawla2005toward}.
\emph{Homomorphic encryption}, on the other hand, can be used to perform computation directly on the encrypted data, thus eliminating the need of exposing any sensitive information~\cite{rivest1978data, gentry2009fully, tebaa2012homomorphic, naehrig2011can}; this technique has been applied, for example, to vehicle positioning data~\cite{popa2009vpriv} and medical records~\cite{molina2009hiccups}.

The flows of data---creation, copying, sharing---can be restricted.
\emph{Information Flow Control} solutions such as~\cite{zdancewic2002programming, sfaxi2010information, zeldovich2008securing} attempt to regulate the flow of information in digital systems. 
\emph{Auditing} implementations such as~\cite{mundada2011silverline, pappas2012cloudfence, ganjali2012auditing} track the data flow by generating usage logs. 
\emph{Data Expiration} makes data inaccessible after a specific time, for example by self-destruction or by invalidating encryption keys~\cite{boneh1996revocable, perlman2005ephemerizer, perlman2005file, geambasu2009vanish}.
\emph{Watermarking} identifies records using hidden fingerprints, to allow traceability and identify leaks~\cite{agrawal2003watermarking, cox2000watermarking, cox1998some}.

\section{Motivation} \label{sec:motivation}

Here we describe our primary motivations for deploying the Copenhagen Networks Study, featuring deep and high-resolution data and a longitudinal approach.

\subsection{Multiplexity}

A majority of big data studies have been done with datasets containing data from a single source, such as call detail records (CDRs)~\cite{onnela_structure_2007}, RFID sensors~\cite{10.1371/journal.pone.0011596}, Bluetooth scanners~\cite{larsen2013crowds}, or online social networks activity~\cite{aral2012identifying}.
Although, as we presented in the Related Work section, analyzing these datasets has led to some exciting findings, we may however not understand how much bias is introduced in such single-channel approaches, particularly in the case of highly interconnected data such as social networks.

We recognize two primary concerns related to the single-source approach: incomplete data and limitation with respect to an interdisciplinary approach.
For social networks, we intuitively understand that people communicate on multiple channels, calling each other, meeting face-to-face, or exchanging emails.
Observing only one channel may introduce bias that is difficult to estimate~\cite{madrigal2013dark}.
Ranjan et al.~investigated in~\cite{Ranjan2012CDR} how CDR datasets---containing samples dependent upon user activity and requiring user participation---may bias our understanding of human mobility.
The authors used data activities as ground truth; due to applications running in the background, sending and requesting data, smartphones exchange data with the network much more often than typical users make calls and without the need for their participation.
Comparing the number of locations and significant locations~\cite{isaacman2011identifying}, they found that the CDRs reveal only a small fraction of users' mobility, when compared to data activity.
The identified home and work locations---considered the most important ones---did not, however, differ significantly when estimated using either of the three channels (voice, SMS, and data).

Domains of science operate primarily on different types of data. Across the sciences, researchers are interested in distinct questions and use very different methods.
Similarly, datasets are obtained from different populations and in different moments, it is difficult to cross-validate or combine findings, and the single-channel origin of the data can be a preventive factor in applying expertise from multiple domains.
If we collect data from multiple channels in the same studies, on the same population, we can work together across field boundaries, utilizing the varied expertises and results they generate to provide more robust insights.

Social networks are `multiplex' in the sense that many different types of links may connect any pair of nodes.
While recent work~\cite{JPScienceCommunitiesTimeMultiplex2010,szell2010mol} begin to explore the topic, a coherent theory describing multiplex, weighted, and directed networks remains beyond the frontier of our current understanding.

\subsection{Sampling}

In many big data studies, data sampling is uneven.
CDRs, for example, only provide data when users actively engage, by making or receiving a phone call or SMS.
Users can also have different patterns of engagement with social networks, some checking and interacting several times a day, while others only once a week~\cite{madden2013teens}.
Further, CDRs are typically provided by a single provider with some finite market share. 
If the market share is $20\%$ of the population and you consider only links internal to your dataset, that translates to only 4\% of the total number of links, assuming random network and random sampling~\cite{onnela_structure_2007}. 
Thus, while CDRs might be sufficient for analyses of mobility, it is not clear that CDRs are a useful basis for social network analysis.
Such uneven, sparse sampling decreases the resolution of data available for analysis.
Ensuring highest possible quality of the data and even sampling is possible with primarily passive data gathering, focusing on digital traces left by participants as they go through their lives, for example by using phones automatically measuring Bluetooth proximity, recording location, and visible WiFi networks~\cite{lazer2009life, aharony2011social, eagle2006reality}.
In cases where we cannot observe users passively or when something simply goes wrong with the data collection, we aim to use the redundancy in the channels: if the user turns off Bluetooth for some period, we can still estimate the proximity of users using WiFi scans (as described in the Results section).

Uneven sampling not only reduces the quality of available data, but also---maybe more importantly---may lead to selection bias when choosing users to include in the analysis.
As investigated in~\cite{Ranjan2012CDR}, when only high-frequency voice-callers are chosen from a CDR dataset for the purpose of analysis, this can incur biases in Shannon entropy values (measure of uncertainty) of mobility, causing overestimation of the randomness of participants' behavior.
Similarly, as shown in~\cite{madden2013teens}, choosing users with a large network and many interactions on Facebook may lead to overestimation of diversity in the ego-networks.
Every time we have to discard a significant number of users, we risk introducing the bias in the data.
Highly uneven sampling that cannot be corrected with redundant data, compels the researcher to make mostly arbitrary choices as part of the analysis, complicating subsequent analysis, especially when no well-established ground truth is available to understand the bias.
Our goal here is to collect evenly sampled high-quality data for all the participants so we do not have to discard anyone; an impossible goal but one worth pursuing.

Since we only record data from a finite number of users, our study population is also a subset, and every network we analyze will be sampled in some way, see~\cite{kossinets2006effects} for a review on sampling.
While the 2013 deployment produces a dataset that is nearly complete in terms of communication between the participants, it is clear that it is subject to other sampling related issues. For example, a relatively small network embedded in a larger society has a large `surface' of links pointing to the outside world, creating a \emph{boundary specification problem}~\cite{laumann:1983}.

\subsection{Dynamics}
The networks and behaviors we observe are not static, displaying dynamics on multiple time-scales.
Long-term dynamics may be lost in big data studies when the participants are not followed for a sufficiently long period and only a relatively narrow slice of data is acquired.
Short-term dynamics may be missed when the sampling frequency is too low.

It is a well established fact that social networks evolve over time~\cite{kossinets2006empirical,Saramäki06012014}.
The time scale of the changes varies and depends on many factors: semester cycle in students life, changing schools or work, or simply getting older, to name just a few.
Without following such dynamics, focusing on a single temporal slice, we risk missing an important aspect of human nature.
To capture it, we need long-term studies, following the participants through the period of months and years. 

Our behavior is not static in even very short intervals.
We have daily routines, meeting with different people in the morning and hanging out with other in the evening, see Figure~\ref{fig:network_dynamics}.
Our workdays may see us going to places and interacting with people differently than on weekends.
It is easy to miss such dynamics when the quality of the data is insufficient, either not sampled frequently enough or of poor resolution, requiring large time bins.

Because each node has a limited bandwidth, only a small fraction of the network is actually `on' at any given time, even if the underlying social network is very dense.
Thus, to get from node A to node B, a piece of information may only travel on links that are active at subsequent times.
Some progress has been made on the understanding of dynamic networks, for a recent review see~\cite{holme:2011}, but in order to understand the dynamics of our highly dense, multiplex network, we need to expand and adapt the current methodologies, for example by adapting the link-based viewpoint to dynamical systems.

\subsection{Feedback}

In many studies, the data collection phase is separated from the analysis.
The data might have been collected during usual operation, before the idea of the study had even been conceived (e.g.~CDRs, Wifi logs), or the access to the data might have not been granted before a single frozen and de-identified dataset was produced.

One real strength of the research proposed here is that, in addition to the richness of the observed data, we are able to run controlled experiments, including surveys distributed via the smartphone software. 
We can for example divide participants into sub-populations and expose them do distinct stimuli, addressing the topic of causality as well as confounding factors both of which have proven problematic~\cite{lyons:2011,shalizi:2011} for the current state of the art~\cite{fowler:2008, christakis:2009}.

Simultaneously, we monitor the data quality not only on the most basic level of a user---number of data points, but also by looking at the entire live dataset to understand if the quality of the collected data is sufficient to answer our research questions.
This allows us to see and fix bugs in the data collection software, or learn that certain behaviors of the participants may introduce bias in the data: for example after discovering missing data, some interviewed students reported turning the phones off for the night to preserve battery.
This allowed us to understand that, even if in terms of the raw numbers, we may be missing some hours of data per day for those users, there was very little information in that particular data anyway.

Building systems with real-time data processing and access allows us to provide the participants with applications and services.
It is an important part of the study not only to collect and analyze the data but also to learn how to create a feedback loop, directly feeding back extracted knowledge on behavior and interactions to the participants.
We are interested in studying how personal data can be used to provide feedback about individual behavior and promote self-awareness and positive behavior change, which is an active area of research in Personal Informatics~\cite{li2010stage}.
Applications for participants create value, increase engagement, and may even allow to deploy studies without buying a large number of smartphones to provide to participants.
Our initial approach has included the development and deployment of a mobile app, which provides feedback about personal mobility and social interactions based on personal participant data~\cite{larsen2013qs}. 
Preliminary results from the deployment, participants surveys, and usage logs suggest an interest in such applications, with a subset of participants repeatedly using the mobile app for personal feedback~\cite{cuttone2013mobile}.
It is clear that feedback can potentially influence the study results: awareness of a certain behavior may cause participants to wish to change that behavior. 
We believe, however, that such feedback is unavoidable in any study, and studying the effects of such feedback (in order to account for it), is an active part of our research.

\subsection{New Science}

The ability to record the highly dynamic networks opens up a new, microscopic level of observation for the study of diffusion on the network.
We are now able to study diffusion of behavior, such as expressions of happiness, academic performance, alcohol and other substance abuse, information, as well as real world infectious disease (e.g.~influenza).
Some of these vectors may spread on some types of links but not others. For example, influenza depends on physical proximity for its spread, while information may diffuse on all types of links; with the deep data approach we can study differences and similarities between various types of spread and the interplay between the various communication channels~\cite{rocha:2011,holme:2010a}.

A crucial step when studying the structure and dynamics of networks is to identify communities (densely connected groups of nodes)~\cite{Fortunato201075,gulbahce_art_2008}. 
In social networks, communities roughly correspond to social spheres.
Recently, we pointed out that communities in many real world network display \emph{pervasive overlap}, where each and every node belongs to more than one group~\cite{ahn:2010}.
It is important to underscore that the question of whether or not communities in networks exhibit pervasive overlap has great practical importance.
For example, the patterns of epidemic spreading change and the optimal corresponding societal countermeasures are very different, depending on the details of the network structure. 

Although algorithms which detect disjoint communities have operated successfully since the notion of graph partitioning was introduced in the 1970's~\cite{fiedler_property_1975}, we point out that most networks investigated so far are highly incomplete in multiple senses. 
Moreover, we can use a simple model to show that sampling could cause pervasively overlapping communities to appear disjoint~\cite{bagrow:2011}.
The results reveal a fundamental problem related to working with incomplete data: \emph{Without an accurate model of the structural ordering of the full network, we cannot estimate the implications of working with incomplete data}.
Needless to say, this fact is of particular importance to studies carried out on (thin) slices of data, describing only a single communication channel, or a fraction of nodes using that channel.
By creating a high-quality, high-resolution data set, we are able to form accurate descriptions of the full data set needed to inform a proper theory for incomplete data.
A deeper understanding of sampling is instrumental for unleashing the full potential of data from the billions of mobile phones in use today.

\section{Data Collection}

The Copenhagen Networks Study aims to address the problem of single-modality data by collecting information from a number of sources that can be used to build networks, study social phenomena, and provide context necessary to interpret the findings. 
A series of questionnaires provides information on the socioeconomic background, psychological traces, and well-being of the participants; Facebook data enables us to learn about the presence and activity of subjects in the biggest online social networking platform~\cite{fb_2013}; finally, the smartphones carried by all participants record their location, telecommunication patterns, and face-to-face interactions.
Sensor data is collected with fixed intervals, regardless of the users' activity, and thus the uneven sampling issue, daunting especially CDR-based studies, is mainly overcome.
Finally, the study is performed on the largest and the most dense population to date in this type of studies. 
The physical density of the participants helps to address the problem of missing data, but raises new questions regarding privacy, since missing data about a person can, in many cases, be inferred from existing data of other participants. 
For example, if we know that person $A$, $B$, and $C$ met at a certain location based on the data from person $A$, we do not need social and location data from $B$ and $C$ to know where and with whom they were spending time.

Below we describe the technical challenges and solutions in multi-channel data collection in 2012 and 2013 deployments. 

\subsection{Data Sources}
The data collected in the two studies was obtained from questionnaires, Facebook, mobile sensing, anthropological field study, and the Wifi system on campus.   

\subsubsection{Questionnaires}
In 2012 we deployed a survey containing 95 questions, covering socioeconomic factors, participants' working habits, and Big Five Inventory (BFI) measuring personality traits~\cite{john1999big}.
The questions were presented as a Google Form and participation in the survey was optional.

In 2013 we posed 310 questions per participant, prepared by a group of collaborating public health researchers, psychologists, anthropologists, and economists from the Social Fabric project (see Acknowledgements). The questions in the 2013 deployment included BFI, Rosenberg Self Esteem Scale~\cite{rosenberg1989society}, Narcissism NAR-Q~\cite{back2013narcissistic}, Satisfaction With Life Scale~\cite{diener1985satisfaction}, Rotters Locus of Control Scale~\cite{rotter1966generalized}, UCLA Loneliness scale~\cite{russell1996ucla}, Self-efficacy~\cite{sherer1982self}, Cohen’s perceived stress scale~\cite{cohen1983global}, Major Depression Inventory~\cite{bech2001sensitivity}, The Copenhagen Social Relation Questionnaire~\cite{lund2014content}, and Panas~\cite{watson1988development}, as well as health- and behavior-related questions. 
The questions were presented using a custom built web application, which allowed for full customization and complete control over privacy and handling of the respondents' data.
The questionnaire application is capable of presenting different types of questions, branching depending on user's answers, and saving each individual's progress. 
The application is available as an open source project at~\url{github.com/MIT-Model-Open-Data-and-Identity-System/SensibleDTUData-Apps-Questionaires}.
Participation in the survey was required for taking part in the experiment. 
In order to track and analyze temporal development, the survey (in a slightly modified form) was repeated every semester on all participating students.

\subsubsection{Facebook Data}
For all participants in both the 2012 and 2013 deployment it was optional to authorize data collection from Facebook, and a large majority opted in.
In the 2012 deployment, only the friendship graph was collected every 24 hours, until the original tokens expired.
In the 2013 deployment, data from Facebook was collected as a snapshot, every 24 hours.
The accessed scopes were birthday, education, feed, friend lists, friend requests, friends, groups, hometown, interests, likes, location, political views, religion, statuses, and work.
We used long-lived Facebook access tokens, valid for 60 days, and when the tokens expired participants receive notification on their phones, prompting them to renew the authorizations.
For the academic study purposes, the Facebook data provided rich demographics describing the participants, their structural (friendship graph) and functional (interactions) networks, as well as location updates.

\subsubsection{Sensor Data}
For the data collection from mobile phones we used a modified version of the Funf framework~\cite{aharony2011social} in both deployments.
The data collection app built using the framework runs on Android smartphones, which were handed out to participants (Samsung Galaxy Nexus in 2012 and LG Nexus 4 in 2013).
All the bugfixes and the improvement of the framework are public and available under the OpenSensing github organization at \url{github.com/organizations/OpenSensing}.

In the 2012 deployment, we manually kept track of which phone was used by each student, and identified data using device IMEI numbers, but this created problems when the phones were returned and then handed out to other participants.
Thus, in the 2013 deployment, the phones were registered in the system by the students in an OAuth2 authorization flow initiated from the phone; the data were identified by a token stored on the phone and embedded in the data files.
The sensed data were saved as locally encrypted sqlite3 databases and then uploaded to the server every 2 hours, provided the phone was connected to WiFi.
Each file contained 1 hour of user data from all probes, saved as a single table.
When uploaded, the data was decrypted, extracted, and included in the main study database.

\subsubsection{Qualitative Data}
An anthropological field study was included in the 2013 deployment.
An anthropologist from the Social Fabric project was embedded within a randomly selected group of approximately 60 students (August 2013 -- august 2014).
A field study consists of participant observation within the selected group, collecting qualitative data while simultaneously engaging in the group activities.
The goal is to collect data on various rationales underlying different group formations while at the same time experiencing bodily and emotionally what it was like to be part of these formations~\cite{ellen1984ethnographic}.
The participant observation included all the student activities and courses, including extracurricular activities such as group work, parties, trips, and other social leisure activities.
All participants were informed- and periodically reminded about the role of the anthropologist.

In addition to its central purpose, the anthropological data adds to the multitude of different data channels, deepening the total pool of data.
This proved useful for running and optimizing project in a number of ways.

Firstly, data from qualitative social analysis are useful---in a very practical sense---in terms of acquiring feedback from the participants. 
One of the goals of the project is to provide value to the participants, in addition to providing quantified-self style access to data, we also created public services: a homepage, a Facebook page, and a blog, where news and information about the project can be posted and commented on.
These services are intended to keep the students interested, as well as to make participants aware of the types and amounts of data collected (see Privacy section).
Because of the anthropologist’s real-world engagement with the students, the qualitative feedback contains complex information about participants’ interests and opinions, including what annoyed, humored, or bored them.
This input has been used to improve existing services, such as visualizations (content and visual expression) and for developing ideas for the future services.
In summary, qualitative insights helped in understanding the participants better and, in turn, to maintain and increase participation.

Secondly, the inclusion of qualitative data increased the potential for interdisciplinary work between computer- and social science.
Our central goal is to capture the full richness of social interactions by increasing the number of recorded communication channels. 
Adding a qualitative social network approach, makes it possible to relate the qualitative observations to the quantitative data obtained from the mobile sensing, creating an interdisciplinary space for methods and theory.
We are particularly interested in the relationship between the observations made by the embedded anthropologist and the data recorded using questionnaires and mobile sensing, to answer questions about the elements difficult to capture using our high resolution approach. Similarly, from the perspective of social sciences, we are able to consider what may be captured by incorporating quantitative data from mobile sensing into a qualitative data pool---and what can we learn about social networks using modern sensing technology. 

Finally, these qualitative data can be used to ground the mathematical modeling process.
Certain things are difficult or impossible to infer from quantitative measurements and mathematical models of social networks, particularly in regards to understanding \emph{why} things happen in the network, as computational models tend to focus on \emph{how}.
Questions about relationship-links severing, tight networks dissolving, and who or what caused it to happen, can be very difficult to answer, but still of importance to understand the dynamics of the social network.
By including data concerned with answering \emph{why} in social networks, we add a new level of understanding to the quantitative data.

\subsubsection{Wifi Data}

For the 2012 deployment, between August 2012 and May 2013, we had been granted access to the campus Wifi system logs.
Every 10 minutes the system provided metadata about all devices connected to the wireless access points on campus (access point MAC address and building location), together with the student ID used for authentication.
We collected the data in a de-identified form, removing the student IDs and matching the users with students in our study.
Campus Wifi data was not collected for the 2013 deployment.

\subsection{Backend System}
\label{subsec:backend}

The backend system, used for data collection, storage, and access, was developed separately for the 2012 and 2013 deployments.
The system developed in 2012 was not designed for extensibility, as it focused mostly on testing various solutions and approaches to massive sensor-driven data collection.
Building on this experience, the system for the 2013 deployment was designed and implemented as an extensible framework for data collection, sharing, and analysis.

\subsubsection{The 2012 Deployment}

The system for the 2012 deployment was built as a Django web application.
The data from the participants from the multiple sources, were stored in a CouchDB database.
The informed consent was obtained by presenting a document to the participants after they authenticated with university credentials.
The mobile sensing data was stored in multiple databases inside a single CouchDB instance and made available via an API.
Participants could access their own data, using their university credentials.
Although sufficient for the data collection and research access, the system performance was not adequate for exposing the data for real-time application access, mainly due to the inefficient de-identification scheme and insufficient database structure optimization.

\subsubsection{The 2013 Deployment}

The 2013 system was built as an open Personal Data System (openPDS)~\cite{de2012trusted} in an extensible fashion.
The architecture of the system is depicted in Figure~\ref{fig:openpdsarch} and consisted of three layers: platform, services, applications.
In the platform layer, the components common for multiple services were grouped, involving identity provider and user-facing portal for granting authorizations.
The identity provider was based on OpenID 2.0 standard and enabled single sign-on (SSO) for multiple applications.
The authorizations were realized using OAuth2 and could be used with both web and mobile applications.
Users enroll into studies by accepting informed consent and subsequently authorizing application to submit and access data from the study.
The data storage was implemented using MongoDB.
Users can see the status and change their authorizations on the portal site, the system included an implementation of the Living Informed Consent~\cite{IMM2013-06632}.

\subsection{Deployment Methods}

Organizing studies of this size is a major undertaking.
All parts from planning to execution have to be synchronized and below we share some considerations and our approaches.
While their main purpose was identical, the two deployments differed greatly in size and therefore also in the methods applied for enrolling and engaging the participants.

\subsubsection{SensibleDTU 2012}
In 2012 approximately 1,400 new students were admitted to the university, divided between two main branches of undergraduate programs.
We focused our efforts on the larger branch containing 900 students, subdivided into 15 study lines (majors).
For this deployment we had $\sim 200$ phones available to distribute between the students.  
To achieve maximal coverage and density of the social connections, we decided to only hand out phones to a few selected majors that had sufficient number of students interested in participating in the experiment. 
Directly asking students about their interest in the study was not a good approach, as it could lead to biased estimates and would not scale well for a large number of individuals.
Instead we appealed to the competitive element of the human nature by staging a competition, running for two weeks from the start of the semester. 
All students had access to a web forum, separate for each major, where they could post ideas that could be realized by the data we would collect, and subsequently vote on their own ideas or three seed ideas, which we provided.
The goal of the competition was twofold, first we wanted students to register with their Facebook account enabling us to study their online social network, and second we wanted to see which major could gain most support (percentage of active students) behind a single idea.
Students were informed about the project and competition by the Dean in person and at one of 15 talks given---one to each major. 
Students were told that our choice of participants would be based on the support each major could muster behind their strongest idea before a given deadline.
This resulted in 24 new research ideas and 1\,026 unique votes.
Four majors gained $>$93\% support behind at least one idea and were chosen to participate in the experiment. 

The physical handout of the phones was split into four major sessions, in which students from the chosen majors were invited; additional small sessions were arranged for students that were unable to attend the main ones.
At each session participants were introduced to our data collection methods, de-identification schemes, and were presented with the informed consent form.
In addition the participants were instructed to fill out the questionnaire. 
A small symbolic deposit in cash was requested from each student; this served partially as compensation for broken phones, but was mainly intended to make participant take better care of the phones, than if they had received them for free~\cite{shampanier2007zero}.
Upon receiving a phone participants were instructed to install the data collector application.
The configuration on each phone was manually checked when participants were leaving---this was particularly important to ensure high quality of data.

This approach had certain drawbacks; coding and setting up the web fora, manually visiting all majors and introducing them to the project and competition, and organizing the handout sessions required considerable effort and time.
However, certain aspects were facilitated with strong support from the central administration of the university.
A strong disadvantage of the outlined handout process is that phones were handed out 3-4 weeks into the semester, thus missing the very first interactions between students. 

\subsubsection{SensibleDTU 2013}

The 2013 deployment was one order of magnitude larger, with 1\,000 phones to distribute.
Furthermore, our focus shifted to engaging the students as early as possible.
Pamphlets informing prospective undergraduate students about the project were sent out along with the official acceptance letters from the university.
Early-birds who registered online via Facebook using links on the pamphlet were promised phones before the start of their studies.
Students from both branches of undergraduate programs were invited to participate (approximately 1\,500 individuals in total), as we expected an adoption percentage between $30\%$ and $60\%$.
Around 300 phones were handed out to early-birds and an additional 200 were handed out during the first weeks of semester. 
As the adoption rate plateaued, we invited undergraduate students from older years to participate in the project. 

The structure of the physical handout was also modified, the participants were requested to enroll online before receiving the phone---accordingly, the informed consent and the questionnaire were part of the registration.
Again we required a symbolic cash deposit for each phone~\cite{shampanier2007zero}.
We pre-installed custom software on each phone to streamline the handout process; students still had to properly set up the phones (making them Bluetooth-discoverable, turning on WiFi connection, etc.).

For researchers considering similar projects with large scale handouts we provide the following recommendations: Engage the pool of subjects as early as possible and be sure to keep their interest. 
Make it easy for participants to contact you, preferably through media platforms aimed at their specific age group. Establish clear procedures in case of malfunctions.
On a side note, if collecting even a small deposit, when multiplied by a factor of 1\,000, the total can add up to significant amount, and must be handled properly. 

\section{Privacy}

When collecting data of very high resolution, over an extended period, from a large population, it is crucial to properly address the privacy of the participants.
We measure the privacy as a difference between what the user understands and consents to regarding her data and what in fact happens to those data.

We believe that ensuring sufficient privacy for the participants, in large part, is the task of providing them with tools to align the data usage with their understanding.
Such privacy tools must be of two kinds: to inform, making the user understand the situation, and to control, aligning the situation with the user's preferences.
There is a tight loop where these tools interact: as the user grows more informed, she may decide to change the settings, and then verify if the change had the expected result.
By exercising the right to information and control, the user expresses Living Informed Consent as described in~\cite{IMM2013-06632}.

Not all students are interested in privacy, in fact we experienced quite the opposite attitude.
During our current deployments, the questions regarding privacy were rarely asked by the participants, as they tended to accept any terms presented to them without thorough analysis.
It is our---the researchers'---responsibility to make the users more aware and empowered to make the right decisions regarding their privacy: by providing the tools, promoting their usage, and engaging in a dialog about the privacy-related issues.

In the 2012 deployment we used a basic informed consent procedure with an online form accepted by the users, after they authenticated with the university account system.
The accepted form was then stored in a database, together with the username, timestamp, and the full text displayed to the user.
The form itself was a text in Danish, describing the study purpose, parties responsible, and participants' rights and obligations.
The full text is available at~\cite{sensibleconsentda} with English translation available at~\cite{sensibleconsenten}. 

\begin{figure}[!ht]
\begin{center}
	\includegraphics[width=0.6\textwidth]{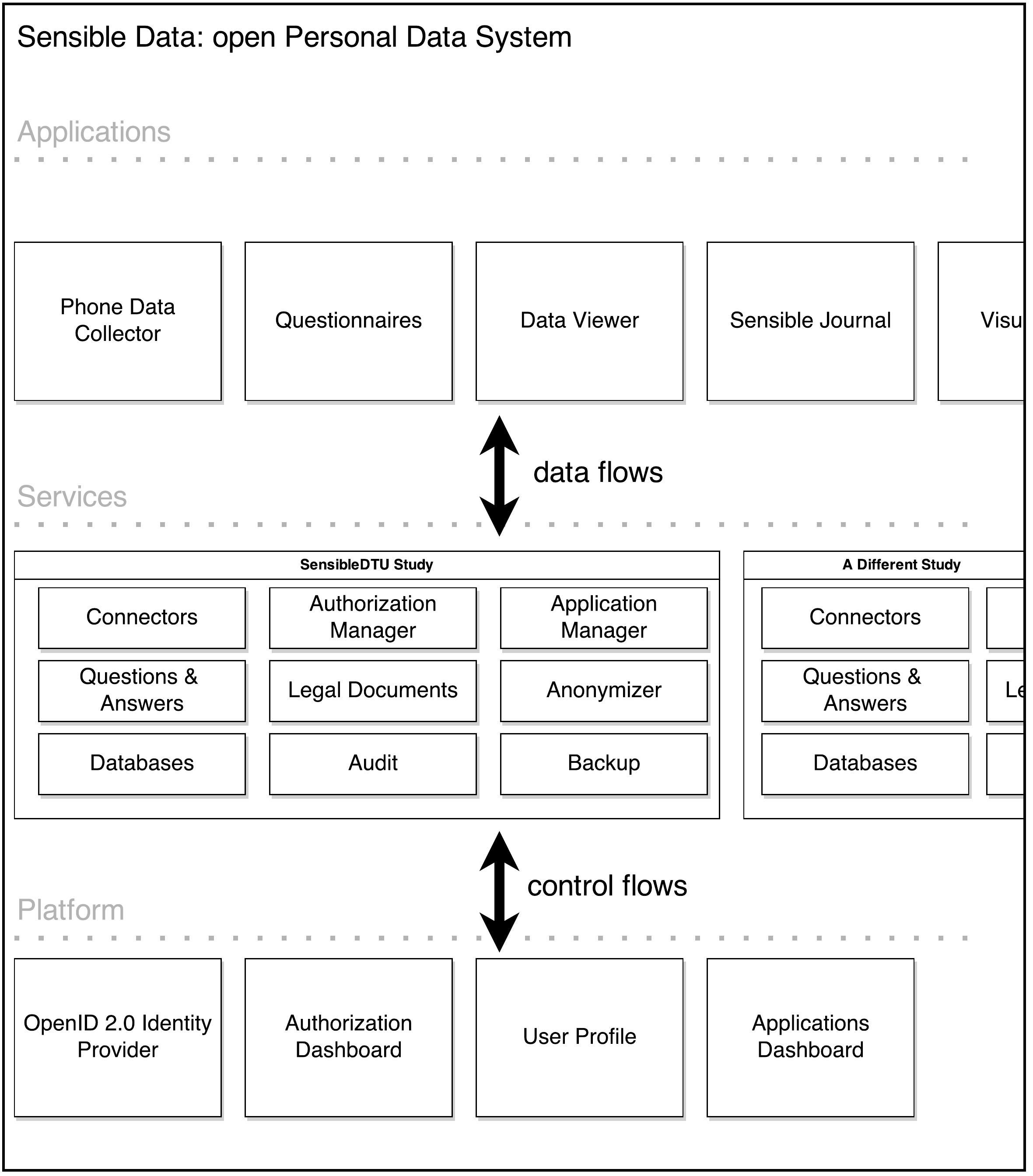}
\end{center}
\caption{
{\bf Sensible Data openPDS architecture.} This system is used in the 2013 deployment and consists of three layers: platform, services, and applications. The platform contains element common for multiple services (in this context: studies). The studies are the deployments of particular data collection efforts. The applications are OAuth2 clients to studies and can submit and access data, based on user authorizations.}
\label{fig:openpdsarch}
\end{figure}

In the 2013 deployment, we used our backend solution (described in Section~\nameref{subsec:backend}) to address the informed consent procedure and privacy in general.
The account system, realized as an OpenID 2.0 server, allowed us to enroll participants, while also supporting research and developer accounts (with different levels of data access).
The sensitive Personally Identifiable Information attributes (PIIs) of the participants were kept completely separate from the user data, all the applications identified users based only on the pseudonym identifiers.
The applications could also access a controlled set of identity attributes for the purpose of personalization (e.g.~greeting the user by name), subject to user OAuth2 authorization.
In the enrollment into the study, after the user had accepted the informed consent document---essentially identical to that from 2012 deployment---a token for a scope \emph{enroll} was created and shared between the platform and service (see Figure~\ref{fig:openpdsarch}).
The acceptance of the document was recorded in the database by storing the username, timestamp, hash of the text presented to the user, as well as the git commit identifying the version of the form.

\begin{figure}[!ht]
\begin{center}
	\includegraphics[width=0.8\textwidth]{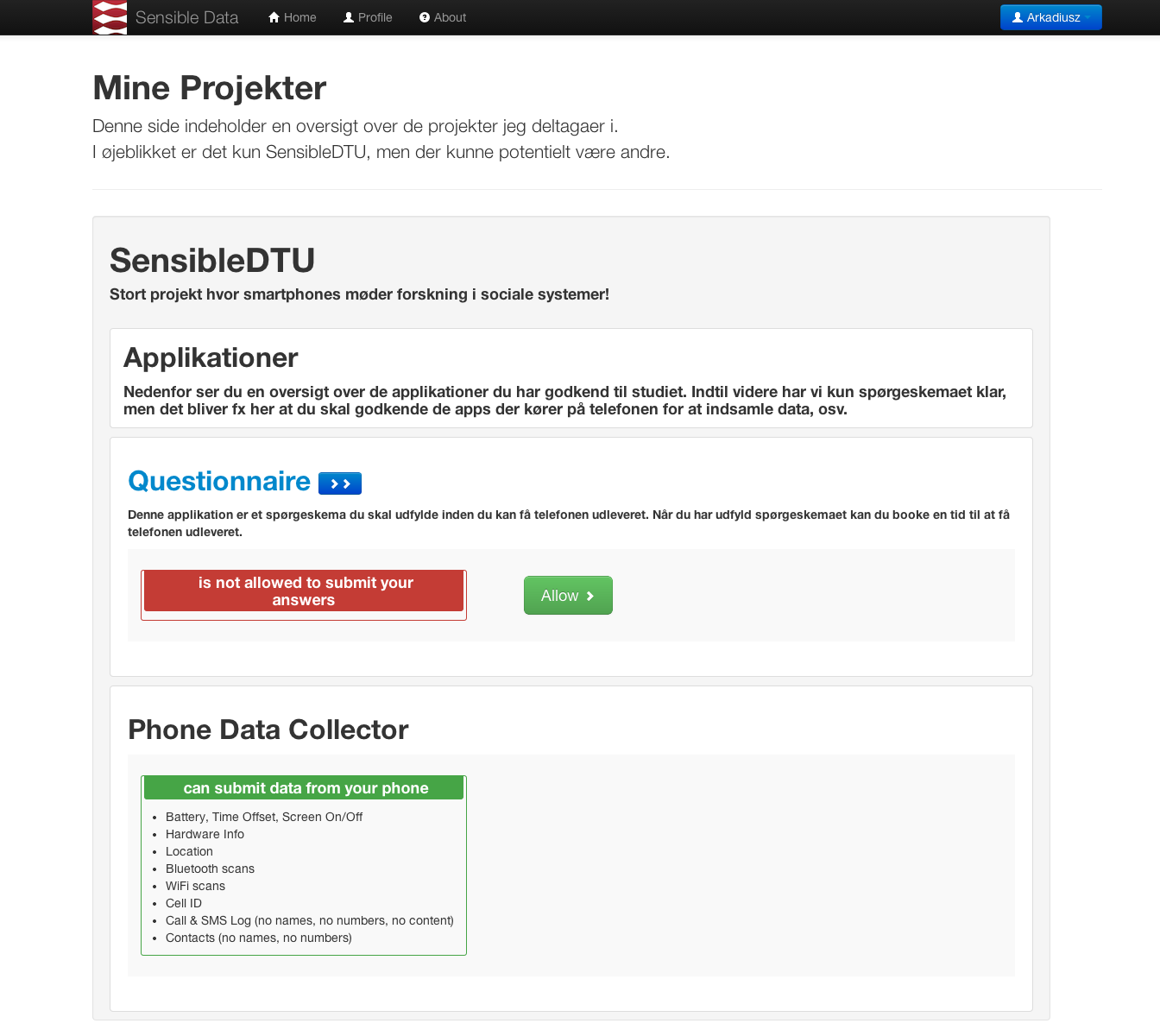}
\end{center}
\caption{
{\bf Authorizations page.} Participants have an overview of the studies they are enrolled into and which applications that are able to submit to and access their data. This is an important step towards users' understanding what is happens with their data and to exercise control over it.}
\label{fig:authorizations}
\end{figure}

All the communication in the system was realized over HTTPS and endpoints were protected with short-lived OAuth2 bearer tokens. 
The text of the documents, including informed consent was stored in a git repository, allowing us to modify anything, while still maintaining the history and being able to reference which version each user has seen and accepted.
A single page overview of the status of the authorizations, presented in Figure~\ref{fig:authorizations}, is an important step in moving beyond lengthy, incomprehensible legal documents accepted by the users blindly and giving more control over permissions to the user.

\begin{figure}[!ht]
\begin{center}
	\includegraphics[width=1\textwidth]{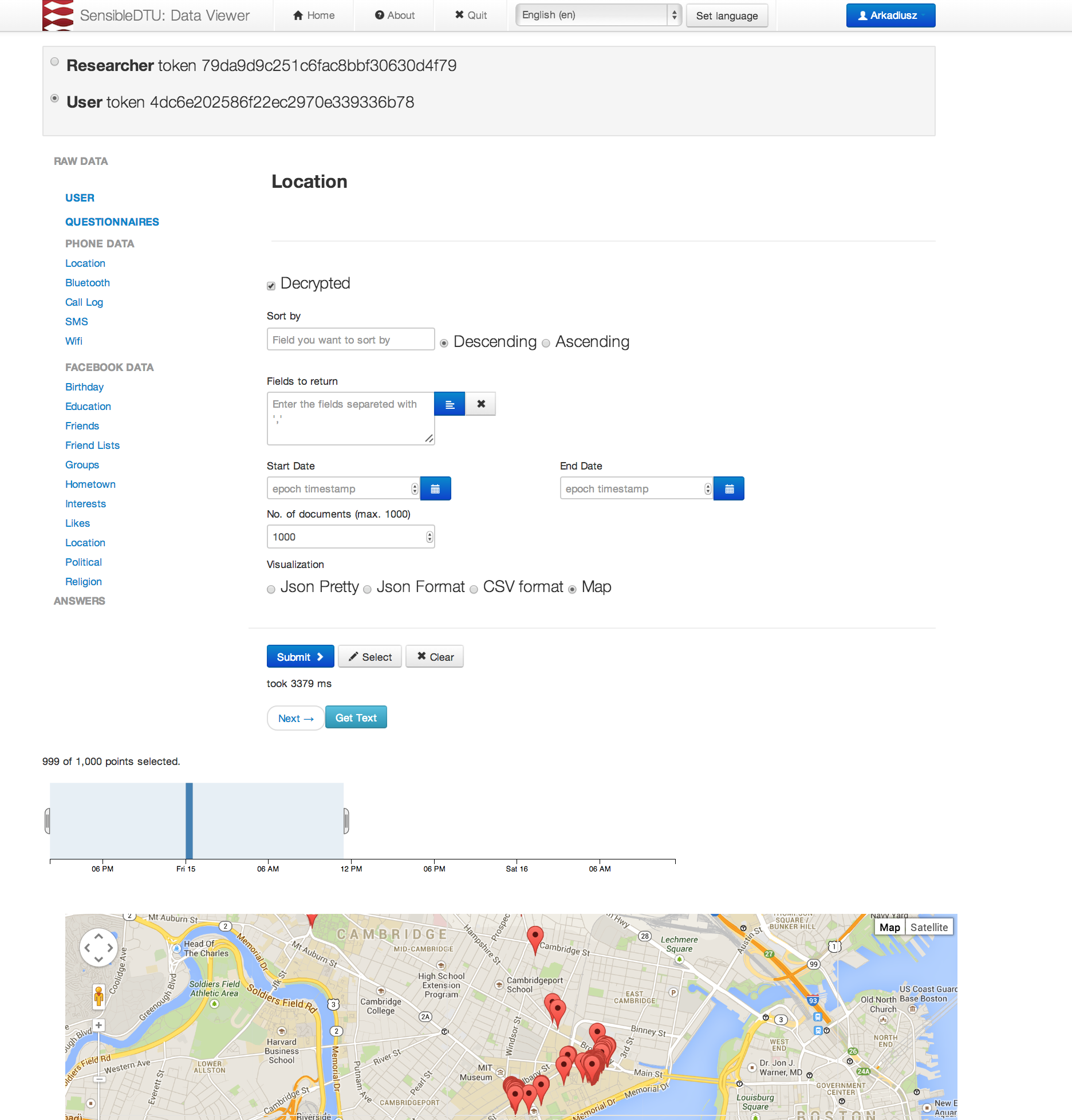}
\end{center}
\caption{
{\bf Data viewer application.} All the collected data can be explored and access via an API. The API is the same for research, application, and end-user access, the endpoints are protected by OAuth2 bearer token.}
\label{fig:fig_data_viewer}
\end{figure}

In the 2013 deployment, the users could access all their data using the same API as the one provided for the researchers and application developers. 
To simplify the navigation, we developed a data viewer application as depicted in Figure~\ref{fig:fig_data_viewer}, which supports building queries with all the basic parameters in a more user-friendly way than constructing API URLs.
Simply having access to all the raw data is however not sufficient, as it is really high-level inferences drawn from the data that are important to understand, for example \textit{Is someone accessing my data to see how fast I drive or to study population mobility?}.
For this purpose we promoted the development of \emph{question \& answer} framework, where the high-level features are extracted from the data before leaving the server, promoting better user understanding of data flows.
This is aligned with the vision of the open Personal Data Store~\cite{de2012trusted}.

Finally, for the purposes of engaging the users in the discussion about privacy, we published blogposts (e.g.~\url{https://www.sensible.dtu.dk/?p=1622}), presented relevant material to students, and answered their questions via the Facebook page(\url{https://www.facebook.com/SensibleDtu}).

\section{Results}



As described in the previous sections, our study has collected comprehensive data about a number of aspects regarding human behavior.
Below, we discuss primary data channels and report some early results and findings.
The results are mainly based on the 2012 deployment due to the availability of longitudinal data.

\subsection{Bluetooth and Social Ties}

Bluetooth is a wireless technology ubiquitous in modern-day mobile devices. 
It is used for short-range communication between devices, including smartphones, handsfree headsets, tablets, and other wearables.
As the transmitters used in mobile devices are primarily of very short range---between 5 and 10 $m$ ($16 - 33$ feet)---detection of the devices of other users (set in `visible' mode) can be used as a proxy for face-to-face interactions~\cite{eagle2006reality}.
We take the individual Bluetooth scans in the form $\left(i,j,t,\sigma \right)$, denoting that device $i$ has observed device $j$ at time $t$ with signal strength $\sigma$.
Bluetooth scans do not constitute a perfect proxy for face-to-face interactions \cite{sekara2014application}, since a) it is possible for people within 10 $m$ radius not to interact socially, and b) it is possible to interact socially over a distance greater than 10 $m$ but, nevertheless, they have been successfully used for sensing social networks~\cite{aharony2011social} or crowd tracking~\cite{stopczynskiparticipatory}.

Between October $1^\text{st}$ 2012 and September $1^\text{st}$ 2013 we collected 12\,623\,599 Bluetooth observations in which we observed 153\,208 unique devices.
The scans on the participants' phones were triggered every five minutes, measured from the last time the phone was powered on. 
Thus, the phones scanned for Bluetooth in a desynchronized fashion, not according to a global schedule.
To account for this, when extracting interactions from the raw Bluetooth scans, we bin them into fixed-length time windows, aggregating the scans within them.
The resulting adjacency matrix, $W_{\Delta t}$ does not have to be strictly symmetric, meaning that user $i$ can observe user $j$ in time-bin $t$, but not the other way around.
Here we assume that Bluetooth scans do not produce false positives (devices are not discovered unless they are really there), and in the subsequent network analysis, we force the matrix to be symmetric, assuming that if user $i$ observed user $j$, the opposite is also true.

\begin{figure}[htbf]
	\includegraphics[width=\hsize]{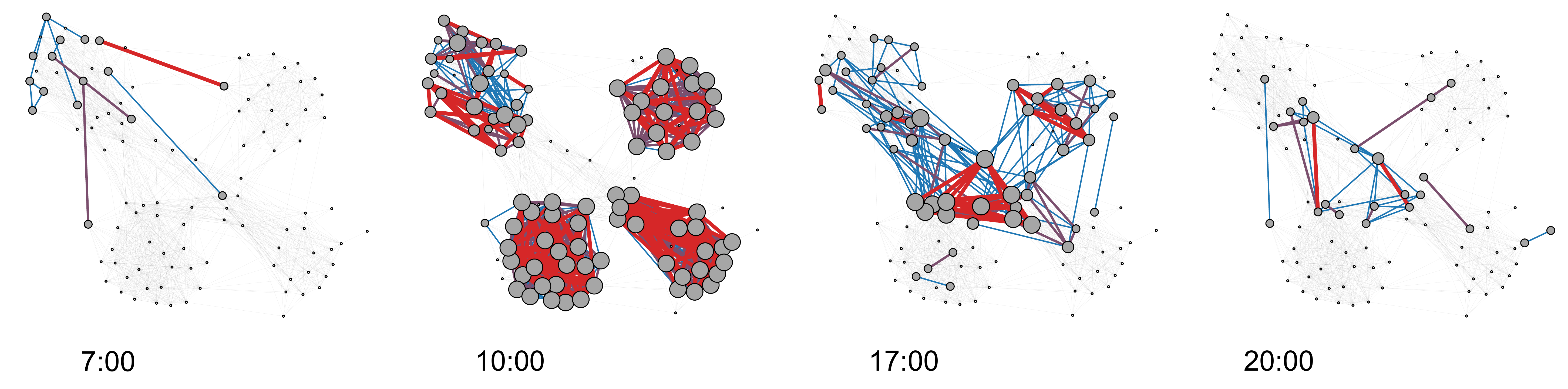}
	\caption{\textbf{Dynamics of face-to-face interactions in the 2012 deployment.} The participants meet in the morning, attend classes within four different study lines, and interact across majors in the evening. Edges are colored according to the frequency of observations, ranging from low (blue) to high (red). Node sizes are scaled according to degree.}
	\label{fig:network_dynamics}
\end{figure}

\begin{figure}[!ht]
\begin{center}
	\includegraphics[width=1\textwidth]{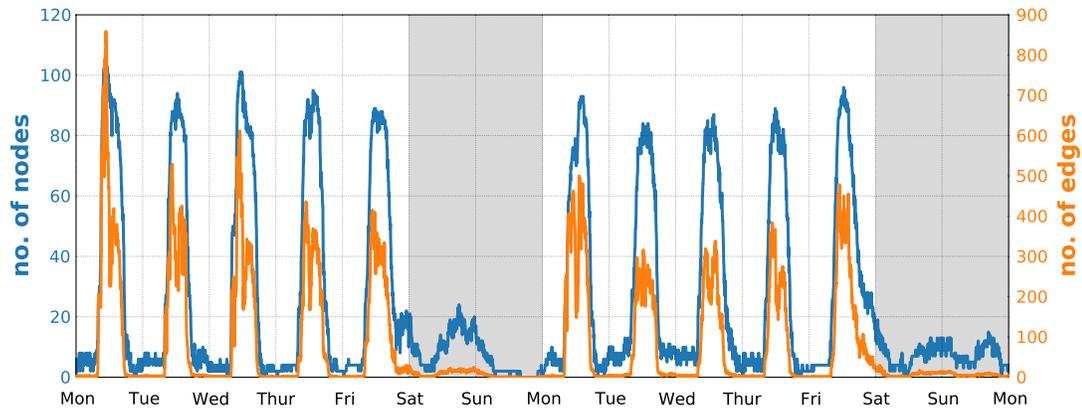}
\end{center}
\caption{
{\bf Weekly temporal dynamics of interactions.} Face-to-face interaction patterns of participants in 5-minute time-bins over two weeks. Only active participants are included, i.e.~users that have either observed another person or themselves been observed in a given time-bin. On average we observed $29$ edges and $12$ nodes in 5-minute time-bins and registered 10\,634 unique links between participants.}
\label{fig:dynamics}
\end{figure}

The interactions between the participants exhibit both daily and weekly rhythms.
Figure~\ref{fig:network_dynamics} shows that the topology of the network of face-to-face meetings changes significantly within single day, revealing academic and social patterns formed by the students.
Similarly, the intensity of the interactions varies during the week, see Figure~\ref{fig:dynamics}.

\begin{figure}[!ht]
	\centering
	\begin{subfigure}[b]{0.45\textwidth}
		\includegraphics[width=\textwidth]{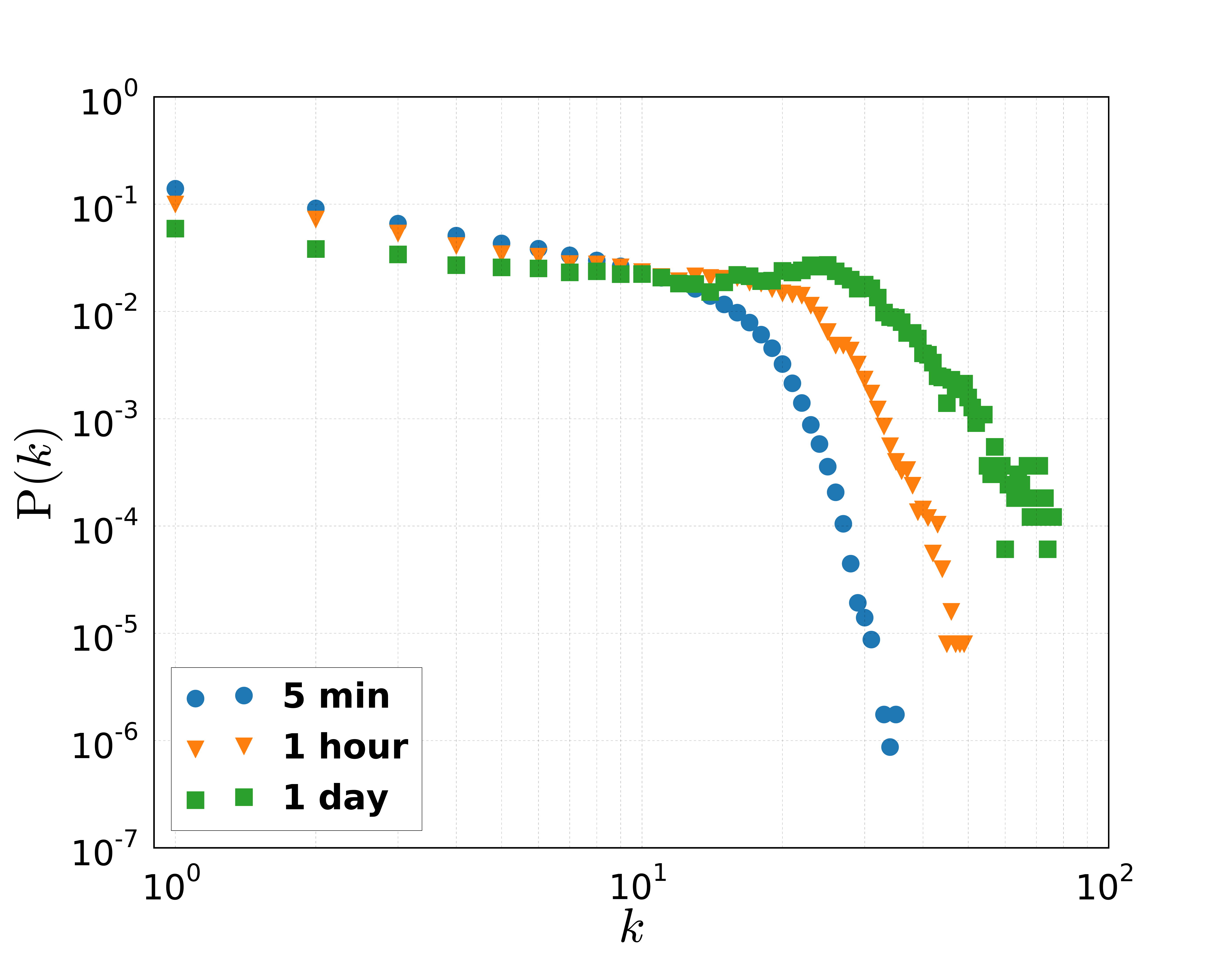}
		\caption{Cumulative degree distributions}\label{fig:bluetooth_degree}
	\end{subfigure}
	\begin{subfigure}[b]{0.45\textwidth}
		\includegraphics[width=\textwidth]{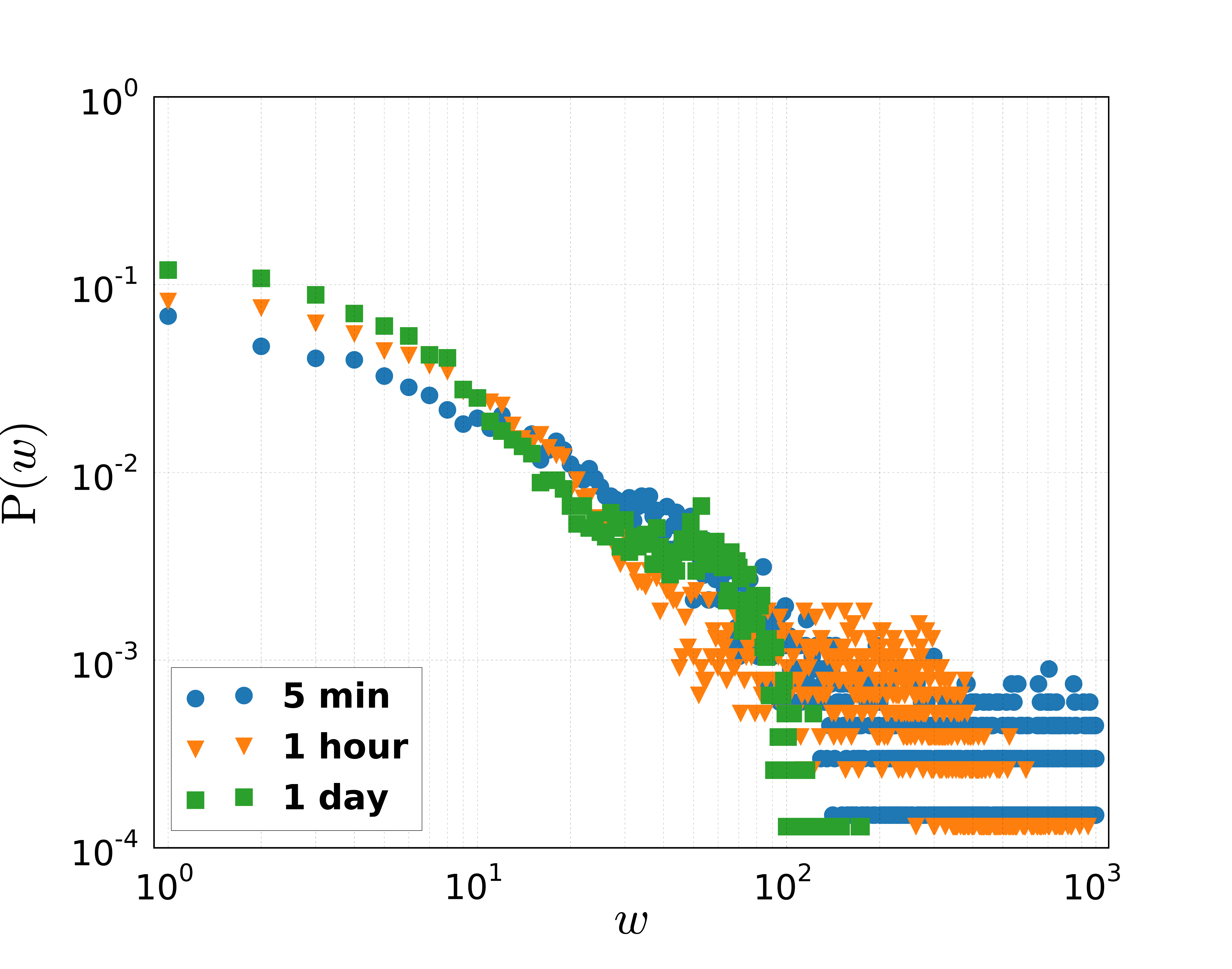}
		\caption{Edge weight distribution}\label{fig:bluetooth_linkweight}
	\end{subfigure}
	\begin{subfigure}[b]{0.45\textwidth}
		\includegraphics[width=\textwidth]{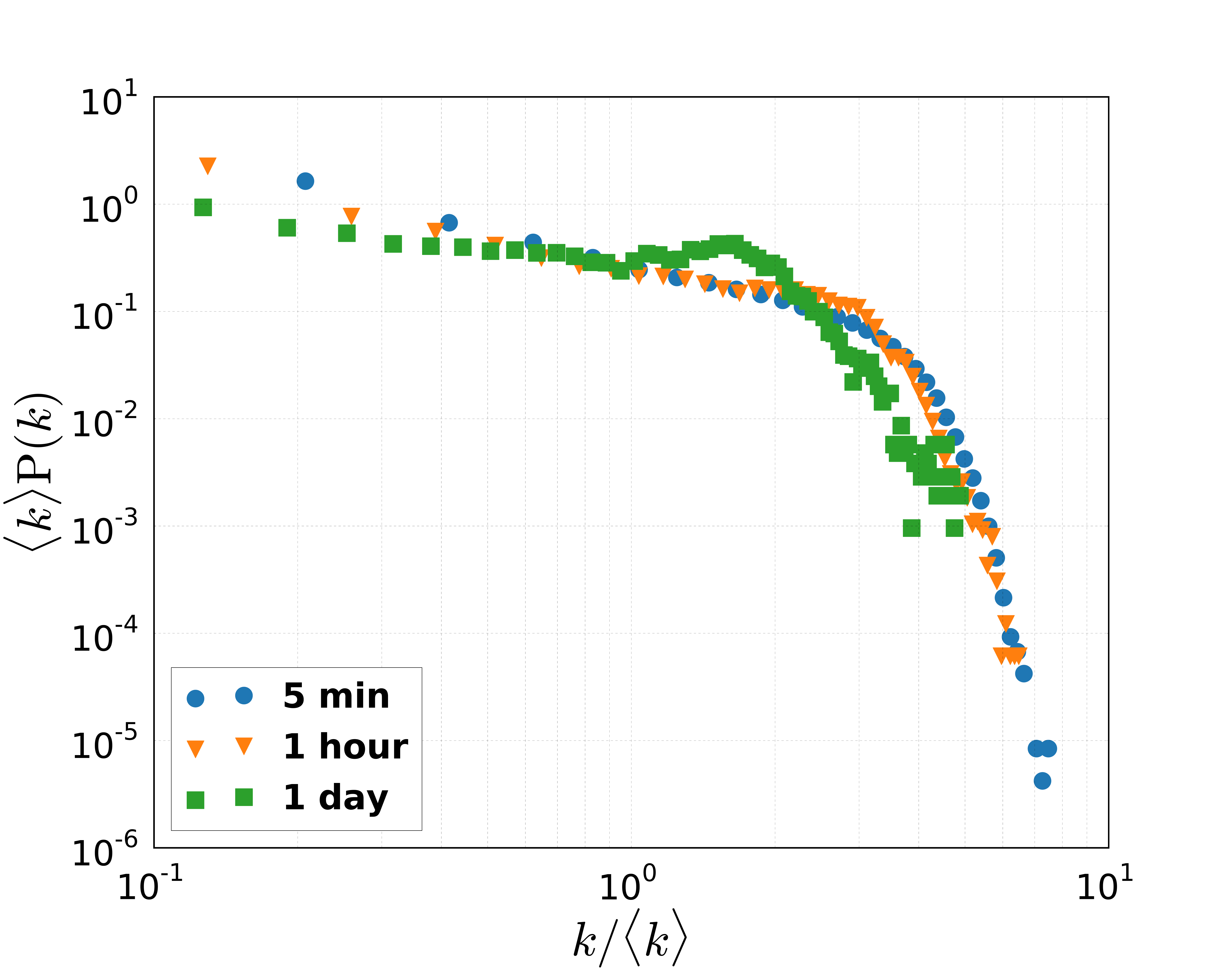}
		\caption{Rescaled degree}\label{fig:bluetooth_degree_rescaled}
	\end{subfigure}
	\begin{subfigure}[b]{0.45\textwidth}
		\includegraphics[width=\textwidth]{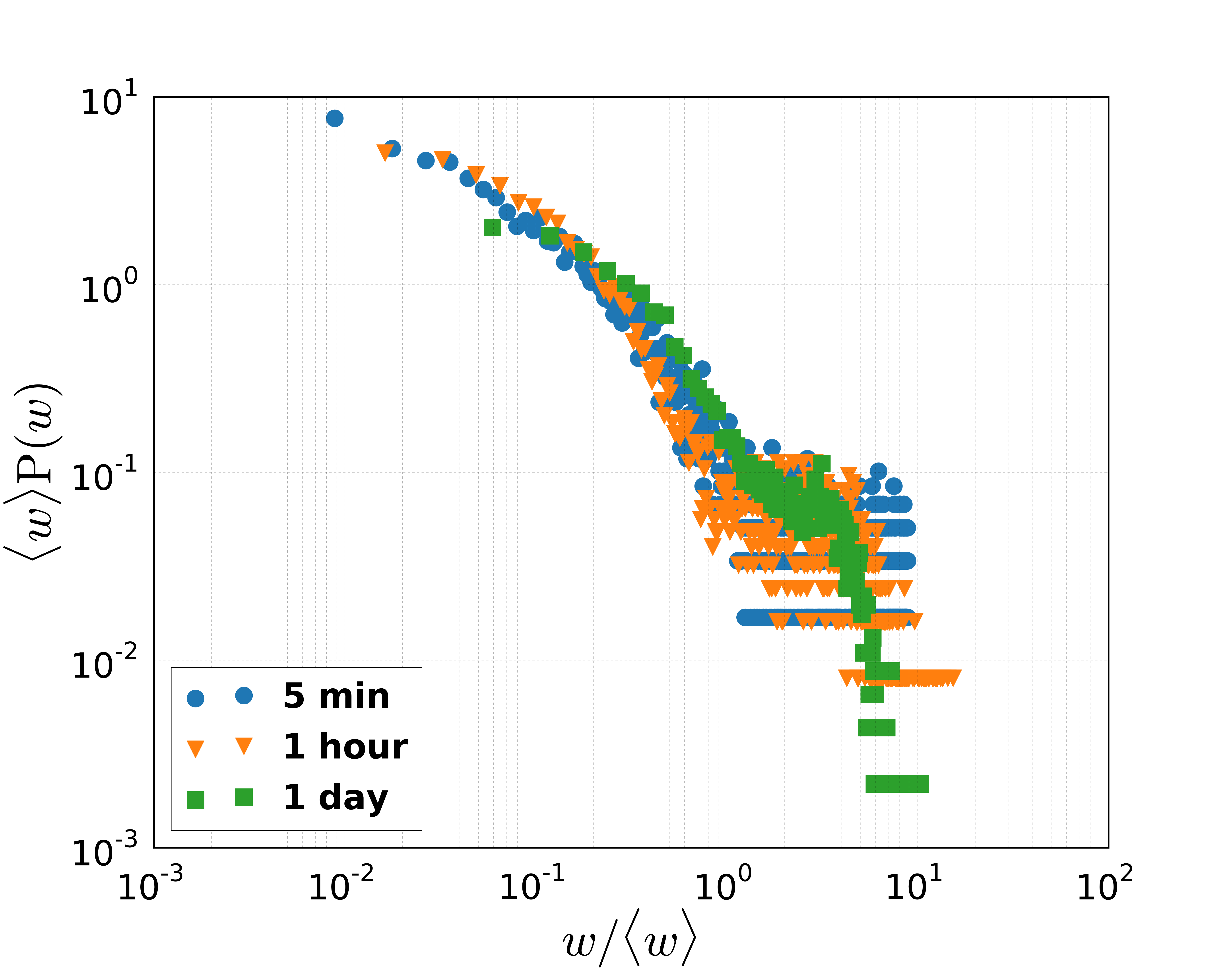}
		\caption{Rescaled edge weight}\label{fig:bluetooth_linkweight_rescaled}
	\end{subfigure}
\caption{{\bf Face-to-face network properties at different resolution levels.} Distributions are calculated by aggregating sub-distributions across temporal window. Differences in rescaled distributions suggest that social dynamics unfold on multiple timescales.}
\label{fig:properties}
\end{figure}

Aggregating over large time-windows blurs the social interactions (network is close to fully connected) while a narrow window reveals detailed temporal structures in the network. 
Figure~\ref{fig:bluetooth_degree} shows the cumulative degree distributions for varying temporal resolutions, with $P(k)$ being shifted towards higher degrees for larger window sizes; this is an expected behavior since each node has more time to amass connections. 
Figure~\ref{fig:bluetooth_linkweight} presents the opposite effect, where the edge weight distributions $P(w)$ shifts towards lower weights for larger windows, this is a consequence on definition of a link for longer time-scales or, conversely, of links appearing in each window on shorter timescales. 
To compare the distribution between timescales, we rescale the properties according to Krings et al.~\cite{krings2012effects} as $P(x)\sim \langle x \rangle P(x/\langle x \rangle)$ with $\langle x \rangle=\sum x P\left(x\right)$ (Figure~\ref{fig:bluetooth_degree_rescaled} and~\ref{fig:bluetooth_linkweight_rescaled}).
The divergence of the rescaled distributions suggest a difference in underlying social dynamics between long and short timescales, an observation supported by recent work on temporal networks~\cite{clauset2012persistence, ribeiro2012quantifying, krings2012effects}.

\subsection{WiFi as Additional Channel for Social Ties}

For the last two decades, wireless technology has transformed our society to the degree where every city in the developed world is now fully covered by mobile~\cite{gsma2012} and wireless networks~\cite{lamarca2005place}.
The data collector application for mobile phones was configured to scan for wireless networks in constant intervals, but also to record the results of scans triggered by any other application running on the phone (`opportunistic' sensing).
Out of the box, Android OS scans for Wifi every 15 seconds, and since we collected these data, our database contains 42\,692\,072 WiFi observations, with 142\,871 unique networks (SSIDs) between October $1^\text{st}$ 2012 and September $1^\text{st}$ 2013 (i.e.~the 2012 deployment).
Below we present preliminary result on Wifi as an additional data-stream for social ties, to provide an example of how our multiple layers of information can complement and enrich each other.

For computational social science using Bluetooth-based detection of participants' devices as a proxy for face-to-face interactions is a well-established method~\cite{aharony2011social, eagle2006reality, miller2012smartphone}. 
The usage of Wifi as a social proxy has been investigated~\cite{kjaergaard2012}, but, to our knowledge, has not yet been used in a large-scale longitudinal study.
For the method we describe here, the users' devices do not sense each other, instead they record the visible beacons (in this instance Wifi access points) in their environment.
Then, physical proximity between two devices---or lack thereof---can be inferred by comparing results of the Wifi scans which occurred within a sufficiently small time window. 
Proximity is assumed if the lists of access points (APs) visible to both devices are similar according to a similarity measure.
We establish the appropriate definition of the similarity measure in a data-driven manner, based on best fit to Bluetooth data.
The strategy is to  compare the lists of results in 10 minute-long time bins, which corresponds to the forced sampling period of Wifi probe as well as to our analysis of Bluetooth data.
If there are multiple scans within the 10 minute bin, the results are compared pair-wise and proximity is assumed if at least one of these comparisons is positive.
The possibility of extracting face-to-face interactions from such signal is interesting, due to the ubiquitous nature of Wifi and high temporal resolution of the signal. 

\begin{figure}[!ht]
	\centering
	\begin{subfigure}[b]{0.45\textwidth}
		\includegraphics[width=\textwidth]{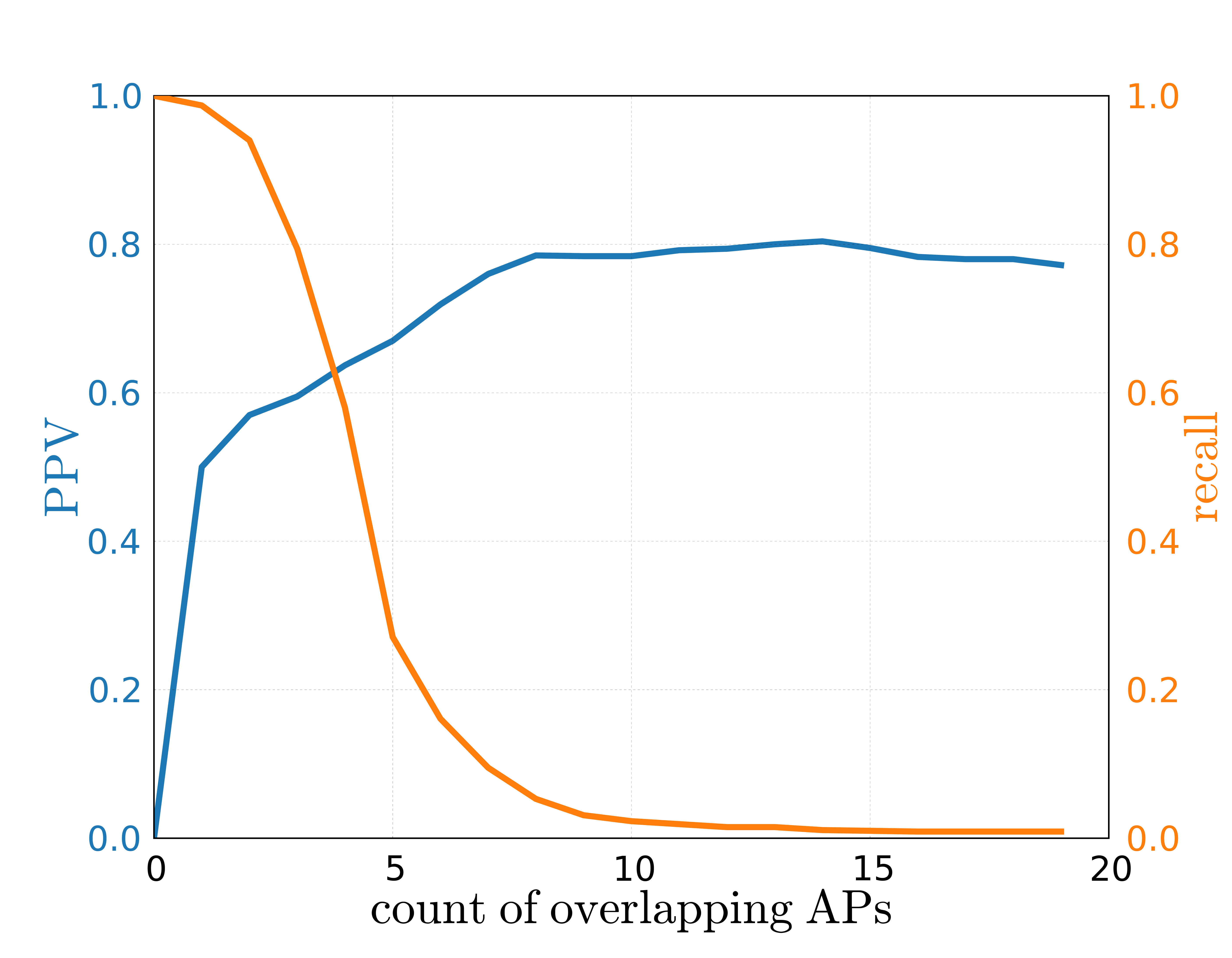}
		\caption{Count of overlapping APs}\label{fig:wifi_measures_count}
	\end{subfigure}
	\begin{subfigure}[b]{0.45\textwidth}
		\includegraphics[width=\textwidth]{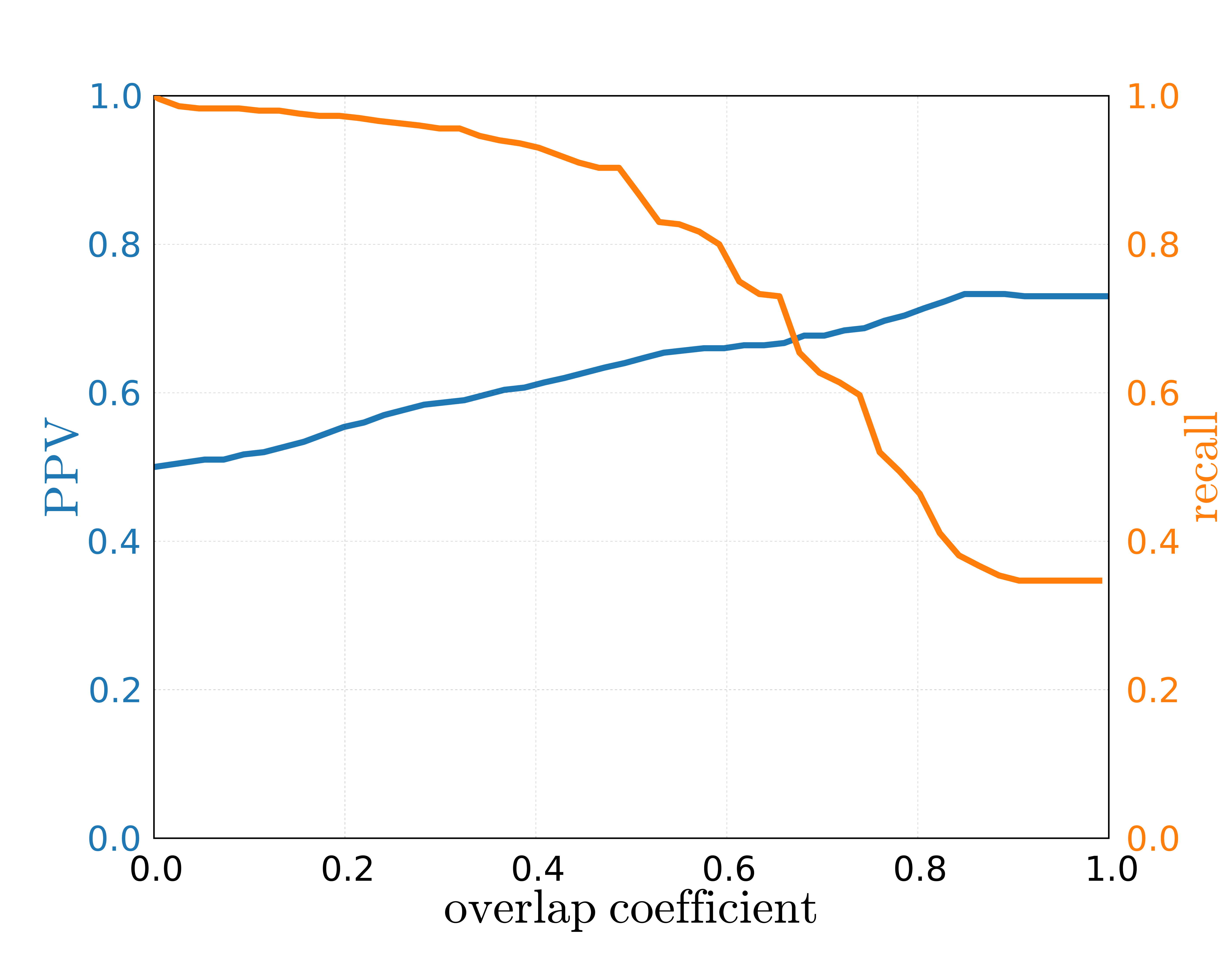}
		\caption{Overlap coefficient}\label{fig:wifi_measures_overlap}
	\end{subfigure}
	\begin{subfigure}[b]{0.45\textwidth}
		\includegraphics[width=\textwidth]{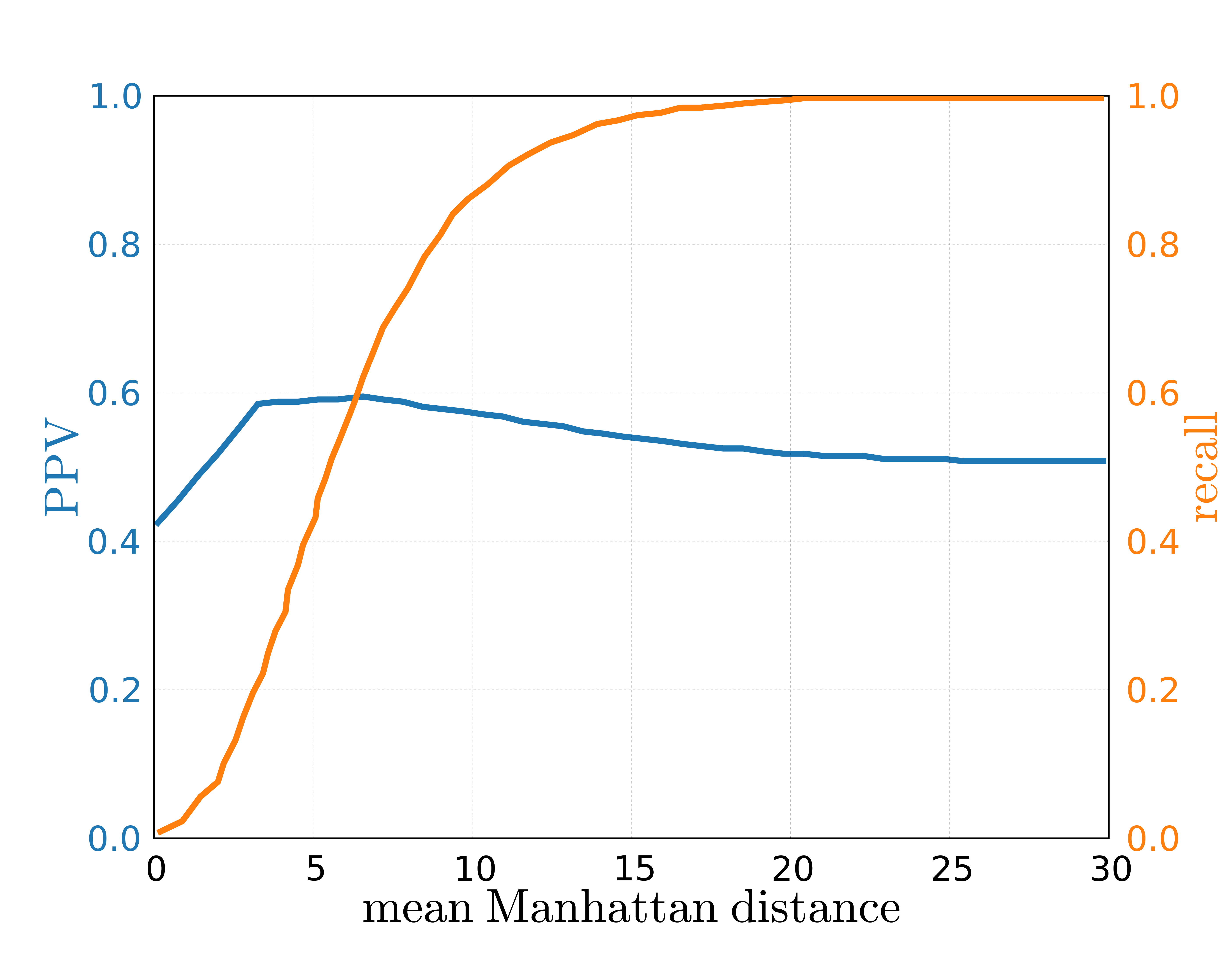}
		\caption{Mean Manhattan distance}\label{fig:wifi_measures_manhattan}
	\end{subfigure}
	\begin{subfigure}[b]{0.45\textwidth}
		\includegraphics[width=\textwidth]{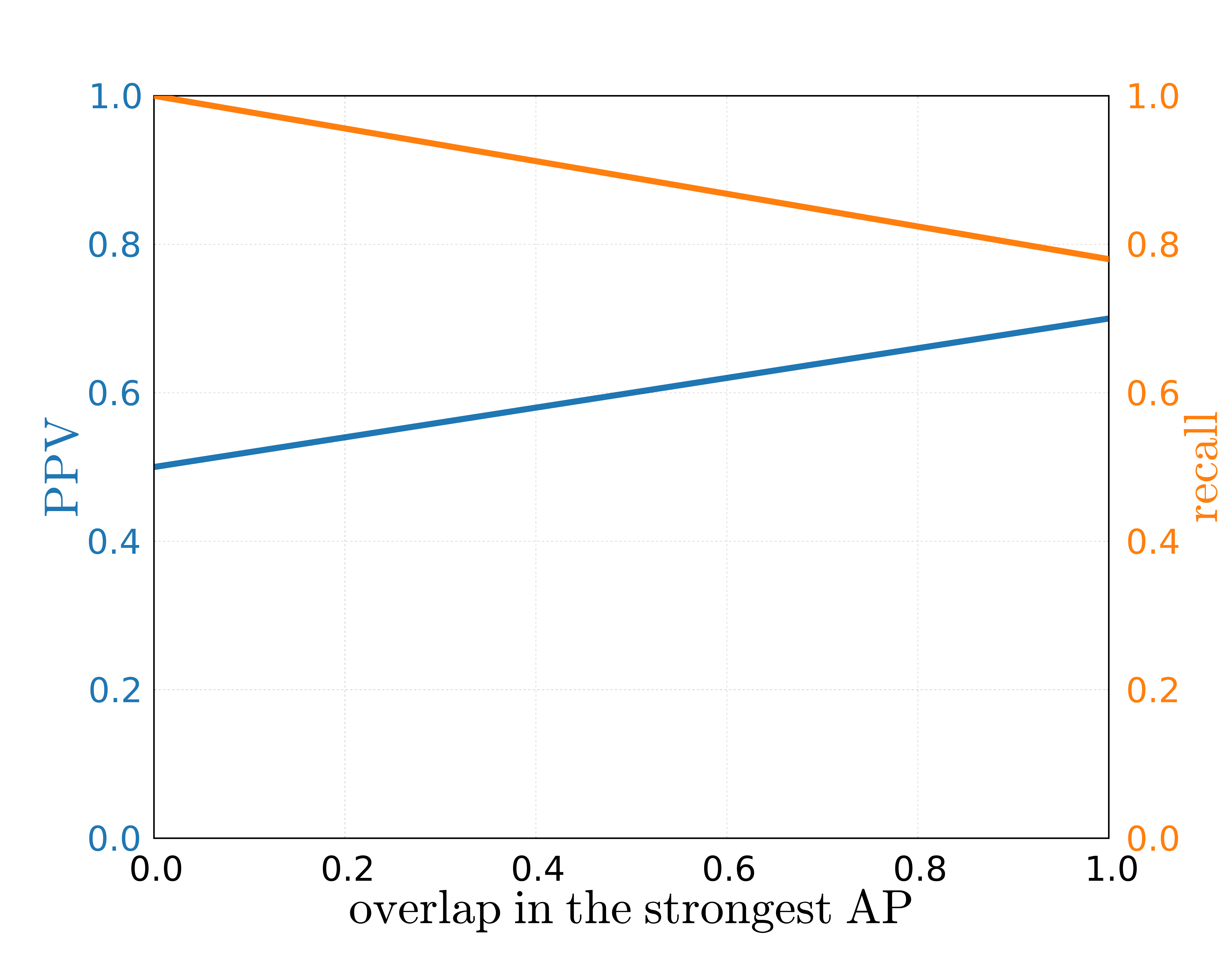}
		\caption{Overlap of the strongest AP}\label{fig:wifi_measures_strongest}
	\end{subfigure}

\caption{\textbf{WiFi similarity measures.} Positive predictive value (precision, ratio of number of true positives and number of positive calls) and recall (sensitivity, fraction of retrieved positives) as functions of parameters in different similarity measures. 
In 98\% of face-to-face meetings derived from Bluetooth, the two devices have also sensed at least one common access point~\textbf{(a)}. 
Identical strongest access point for two separate mobile devices, is a strong indication of a face-to-face meeting~\textbf{(d)}.}
\label{fig:wifi_measures}
\end{figure}

We consider four measures and present their performance in Figure~\ref{fig:wifi_measures}.
Figure~\ref{fig:wifi_measures_count} shows the positive predictive value and recall as a function of minimum number of overlapping access points (${| X \cap Y | }$) required to assume physical proximity.
In approx.~$98\%$ of all Bluetooth encounters, at least one access point was seen by both devices, the recall however drops quickly with the increase of their required number. 
This measure favors interactions in places with high number of access points, where it is more likely that devices will have a large scan overlap. 
The result confirms that lack of a common AP has a very high positive predictive power as a proxy for \emph{lack} of physical proximity, as postulated in~\cite{carlotto2008proximity}.
Note, that for the remaining measures, we assume at last one overlapping AP in the compared lists of scan results.

The overlap coefficient defined as $\mathrm{overlap}(X,Y) = \frac{| X \cap Y | }{\min(|X|,|Y|)}$ penalizes encounters taking place in Wifi-dense areas, due to higher probability of one device picking up a signal from a remote access point which is not available to the other device, see Figure~\ref{fig:wifi_measures_overlap}.

Next, we compare the received signal strengths between overlapping routers using the mean $\ell_{1}$-norm (mean Manhattan distance, $\frac{{|| X \cap Y|| }_1}{|X \cap Y|}$). 
Received signal strenght (RSSI) is measured in $dBm$ and the Manhattan disance between two routers is the difference in the RSSI between them, measured in $dB$.
Thus, mean Manhattan distance is the mean difference in received signal strength of the overlapping routers in the two compared scans.

Finally, we investigate the similarity based on the router with the highest received signal strength---the proximity is assumed whenever it is the same access point for both devices, $\mathrm{max}(X) = \mathrm{max}(Y)$. 
This measure provides both high recall and positive predictive value and, after further investigation for the causes for errors, is a candidate proxy for face-to-face interactions.

The performance of face-to-face event detection based on Wifi can be further improved by applying machine learning approaches~\cite{comm2sense,carlotto2008proximity}. 
It is yet to be established, by using longitudinal data, whether the errors in using single features are caused by inherent noise in measuring the environment, or if there is a bias which could be quantified and mitigated.
Most importantly, the present analysis is a proof-of-concept and further investigation is required to verify if networks inferred from Wifi and Bluetooth signals are satisfyingly similar, before Wifi can be used as an autonomous channel for face-to-face event detection in the context of current and future studies. 
Being able to quantify the performance of multi-channel approximation of face-to-face interaction and to apply it in the data analysis is crucial to address the problem of missing data as well as to estimate the feasibility and understand the limitations of single-channel studies.

\subsection{Location and Mobility}

A number of applications ranging from urban planning, to traffic management, to containment of biological diseases rely on ability to accurately predict human mobility. 
Mining location data allows to extract semantic information such as points of interest, trajectories, and modes of transportation~\cite{lin2013mining}.
In this section we report preliminary results of an exploratory data analysis of location and mobility patterns.

\begin{figure}[!ht]
	\centering
	\begin{subfigure}[t]{0.45\textwidth}
		\includegraphics[width=\textwidth]{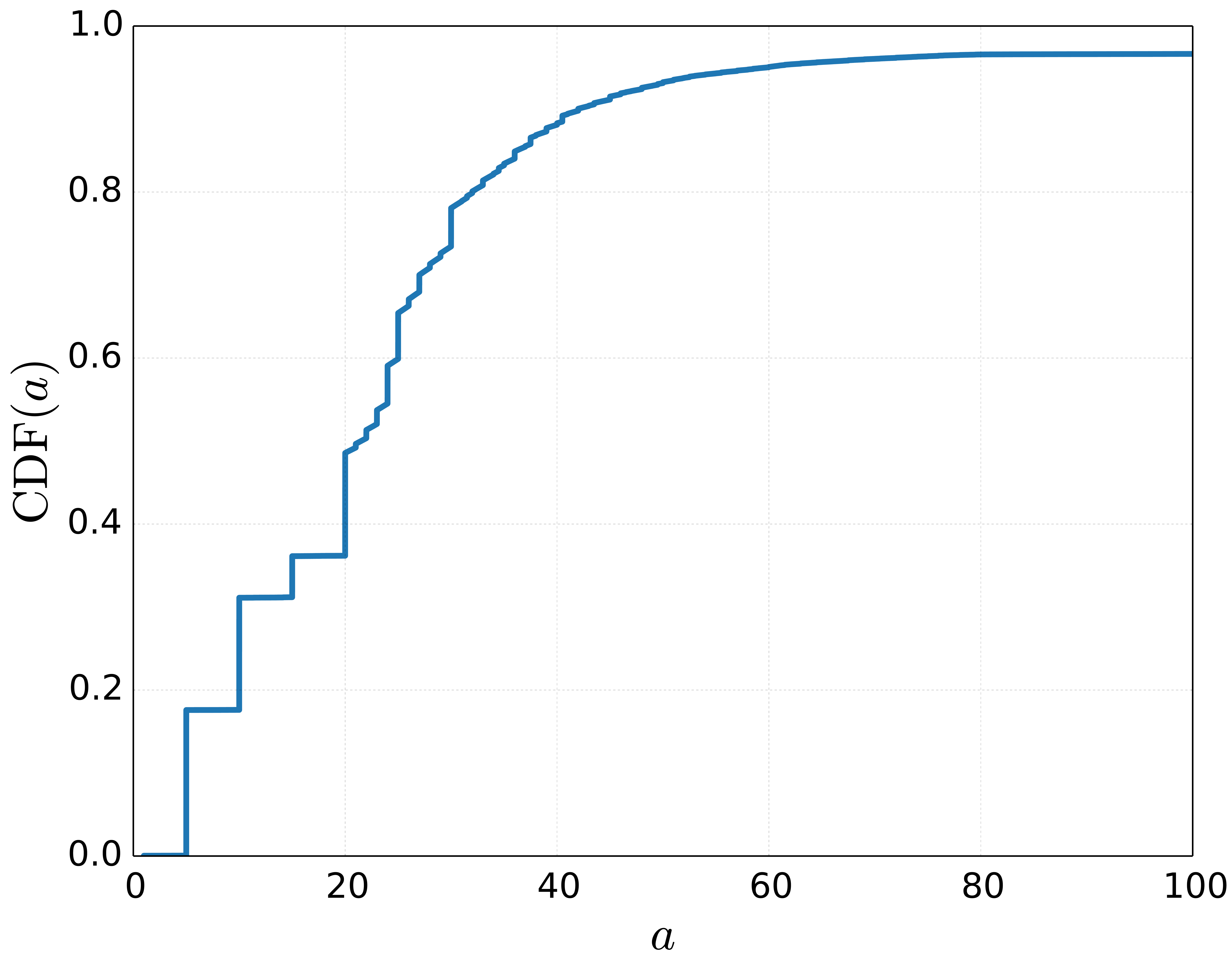}
		\caption{Estimated cumulative distribution function for the accuracy of samples $a$ (in meters).}\label{fig:location1}
	\end{subfigure}
	\begin{subfigure}[t]{0.45\textwidth}
		\includegraphics[width=\textwidth]{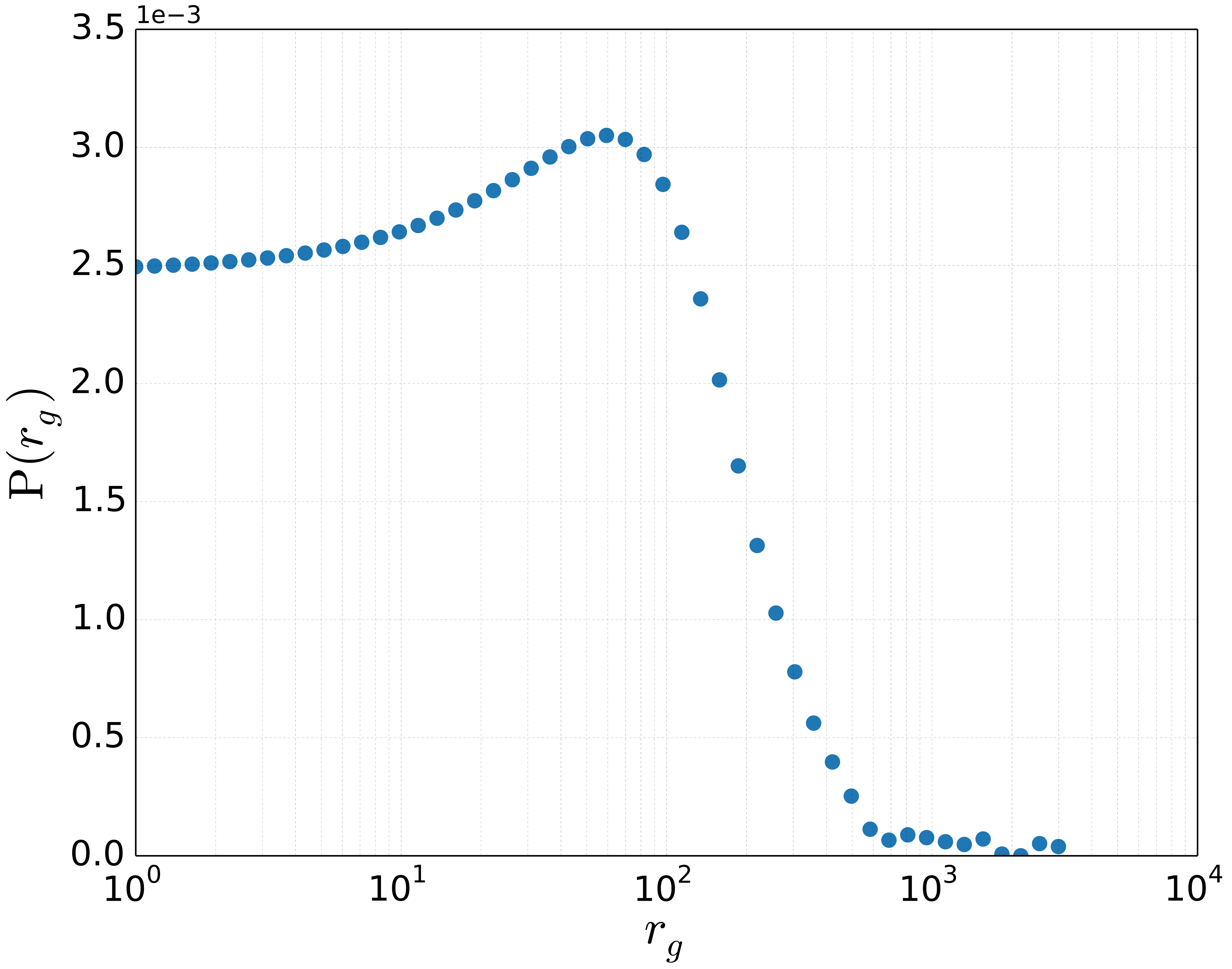}
		\caption{Gaussian kernel density estimation for the radius of gyration $r_g$ (in kilometers).}\label{fig:location2}
	\end{subfigure}
	\begin{subfigure}[t]{0.45\textwidth}
		\includegraphics[width=\textwidth]{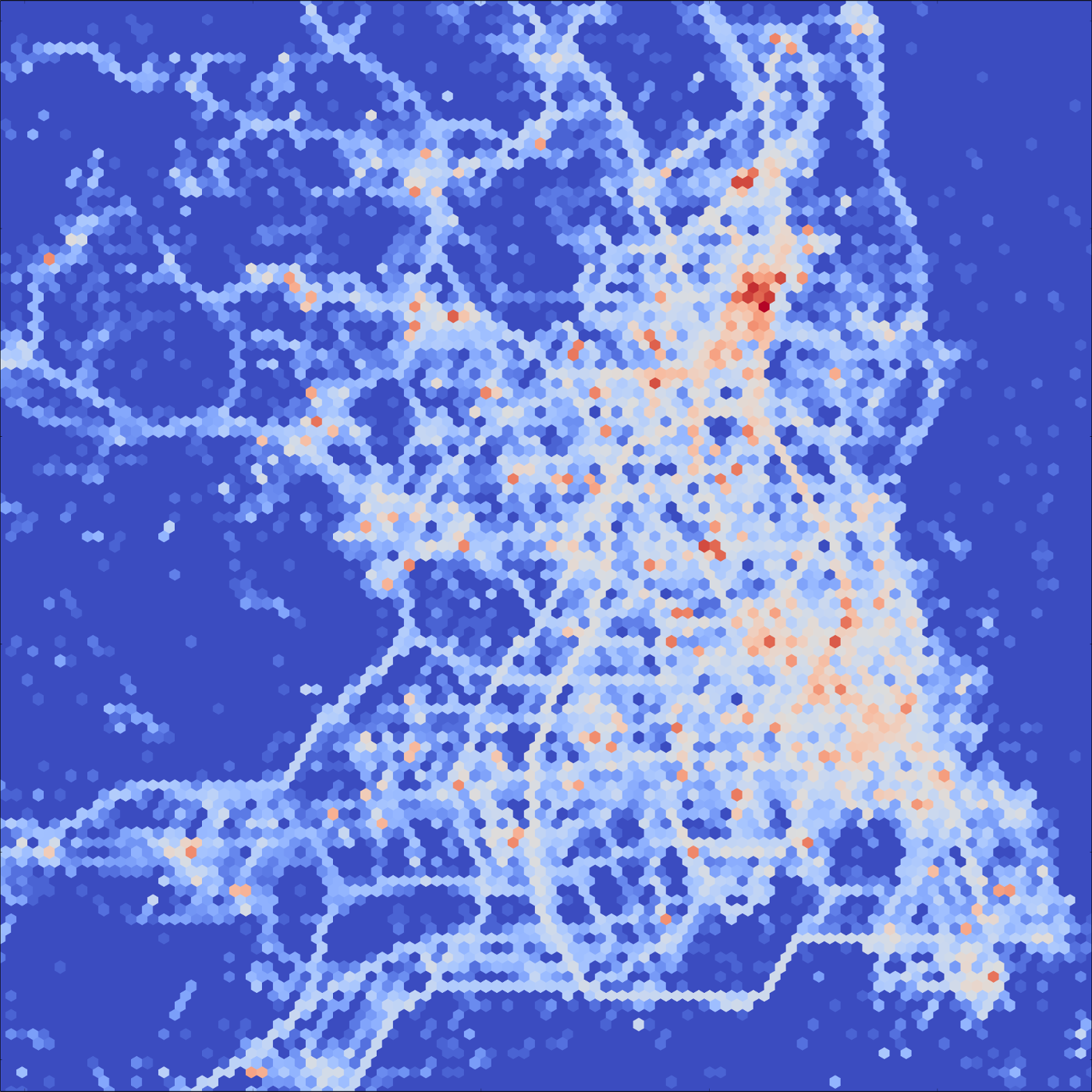}
		\caption{Two-dimensional histogram for the locations, with hex binning and logarithmic color scale.}\label{fig:location3}
	\end{subfigure}
	\begin{subfigure}[t]{0.45\textwidth}
		\includegraphics[width=\textwidth]{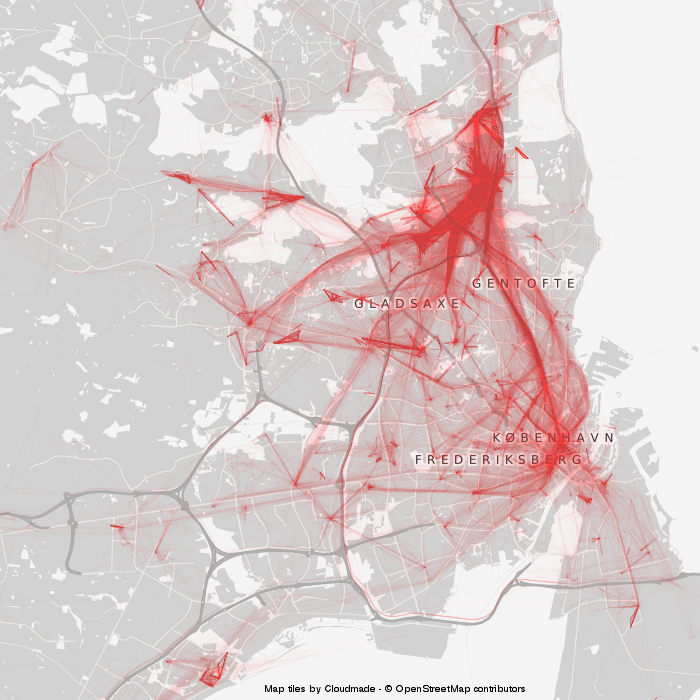}		
		\caption{Cumulated transitions between stop locations.}\label{fig:location4}
	\end{subfigure}

\caption{\textbf{Location and Mobility.} We show the accuracy of the collected samples, radius of gyration of the users, and identify patterns of collective mobility.}
\label{fig:location}
\end{figure}

Location data was obtained by periodically collecting the best position estimate from the location sensor on each phone, as well as recording location updates triggered by other applications running on the phone (opportunistic behavior). 
In total we collected 7\,593\,134 data points in 2012 deployment the form (userid, timestamp, latitude, longitude, accuracy).
The best-effort nature of the data presents new challenges when compared to the majority of location mining literature, which focuses on high-frequency, high-precision GPS data.
Location samples on the smartphones can be generated by different providers, depending on the availability of the Android sensors, as explained in~\url{developer.android.com/guide/topics/location/strategies.html}.
For this reason, accuracy of the collected position can vary between few meters for GPS locations, to hundreds of meters for cell tower location. 
Figure~\ref{fig:location1} shows the estimated cumulative distribution function for the accuracy of samples; almost $90\%$ of the samples have reported accuracy better than 40 meters.

We calculate the radius of gyration $r_g$ as defined in~\cite{song2010limits} and approximate the probability distribution function using a gaussian kernel density estimation, determining the bandwidth value by cross-validation (Figure~\ref{fig:location2}). 
The kernel density peaks around $10^2$ km and then rapidly goes down, displaying a fat-tailed distribution.
Manual inspection of the few users with $r_g$ around $10^3$ km revealed that travels abroad can amount to such high mobility.
Although we acknowledge that this density estimation suffers due to the low number of samples, our measurements suggest that real user mobility is underestimated in studies based solely on CDRs, such as in~\cite{song2010limits}, as they fail to capture travels outside of the covered area.

Figure~\ref{fig:location3} shows a two-dimensional histogram of the locations, with hexagonal binning and logarithmic color scale (from blue to red).
The red hotspots identify the most active places, such as the university campus and dormitories.
The white spots are the frequently visited areas, such as major streets and roads, stations, train lines, and the city center.
 
From the raw location data we can extract stop locations as groups of locations clustered within distance $D$ and time~$T$\cite{hariharan2004project, zheng2009mining, montoliu2010discovering, zheng2010collaborative}.
By drawing edges between stop locations for each user, so that the most frequent transitions stand out, we can reveal patterns of collective mobility (Figure~\ref{fig:location4}).

\subsection{Call \& Text Communication Patterns}
With the advent of mobile phones in the late $20^{\text{th}}$ century, the way we communicate has changed dramatically.
We are no longer restricted to landlines and are able to move around in physical space while communicating over long distances. 

The ability to efficiently map communication networks and mobility patterns (using cell towers) for large populations, has made quantification of human mobility patterns possible, including investigations of social structure evolution~\cite{palla2007quantifying}, economic development~\cite{eagle2010network}, human mobility~\cite{gonzalez2008understanding,song2010limits}, spreading patterns~\cite{wang2009understanding}, and collective behavior with respect to emergencies~\cite{bagrow2011collective}.
In the study, we have collected call logs from each phone as (caller, callee, duration, timestamp, call type), where the call type could be incoming, outgoing, or missed.
Text logs contained (sender, recipient, timestamp, incoming/outgoing, one-way hash of content).

In 2012 deployment we collected 56\,902 incoming and outgoing calls, of which 42\,157 had duration larger than zero.
The average duration of the calls was $\langle d \rangle = 142.04 s$, with the median duration of $48.0 s$.
The average ratio between incoming and outgoing calls for the user was $r_{in/out}=0.98$.
In the same period we collected 161\,591 text messages with the average ratio for user $r_{in/out}=1.96$.

\begin{figure}[!ht]
\begin{center}
	\includegraphics[width=0.6\textwidth]{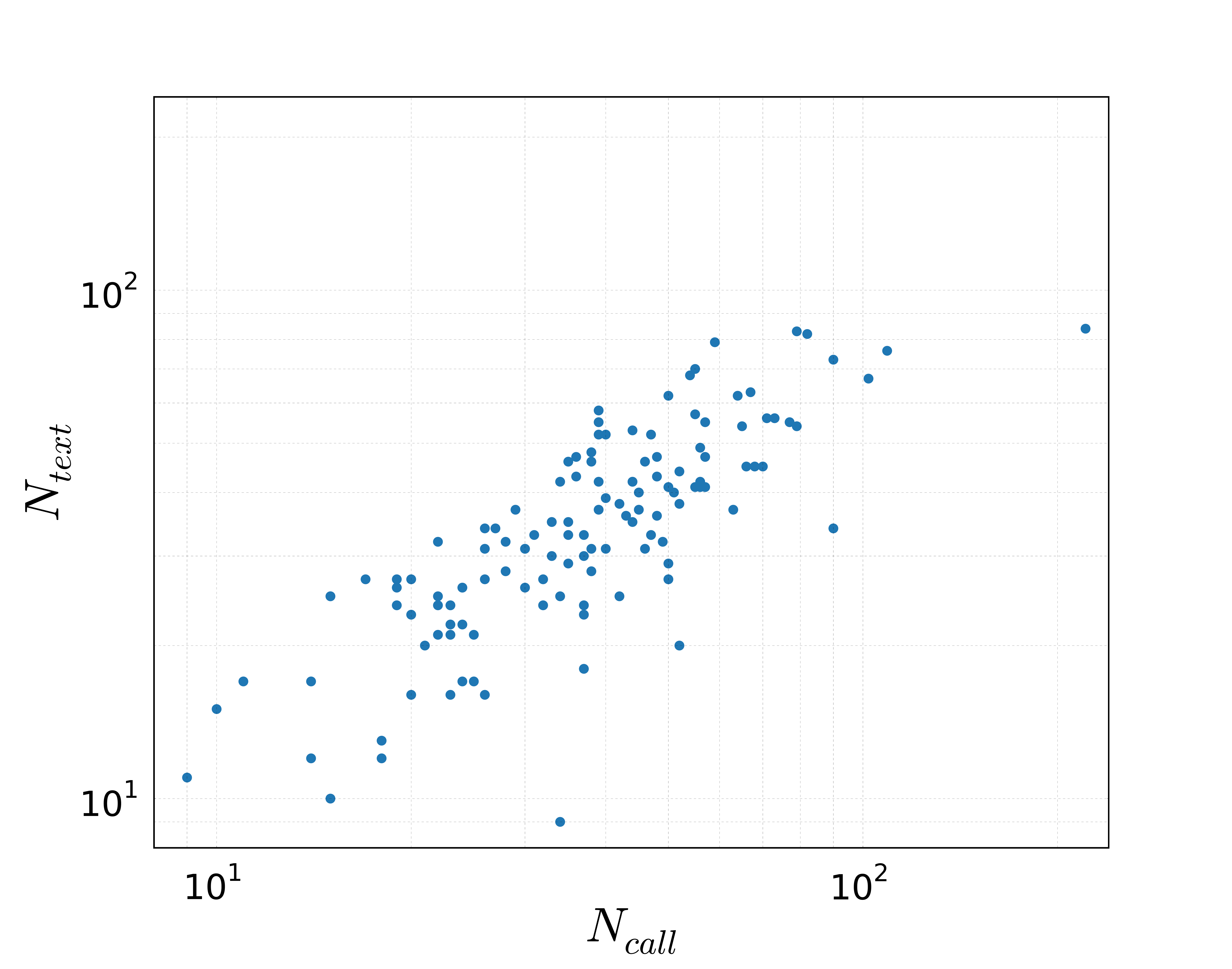}
\end{center}
\caption{
{\bf Diversity of communication logs.} Diversity is estimated as the set of unique numbers that a person has contacted or been contacted by in the given time period on given channel. We note strong correlation in diversity, whereas the similarity of the sets of nodes is fairly low.}
\label{fig:com_logs}
\end{figure}

We find a correlation of $0.75$ between the number of unique contacts users contacted via SMS and voice calls, as depicted in Figure \ref{fig:com_logs}. 
However, the similarity $\sigma=|N_{call}\cap N_{text}|/|N_{call}\cup N_{text}|$ between the persons a participant contacts via calls ($N_{call}$) and SMS ($N_{text}$) is on average $\langle \sigma \rangle=0.37$, suggesting that even though users utilize both forms of communication in similar capacity, those two are, in fact, used for distinct purposes. 

\begin{figure}[!ht]
\begin{center}
	\includegraphics[width=1\textwidth]{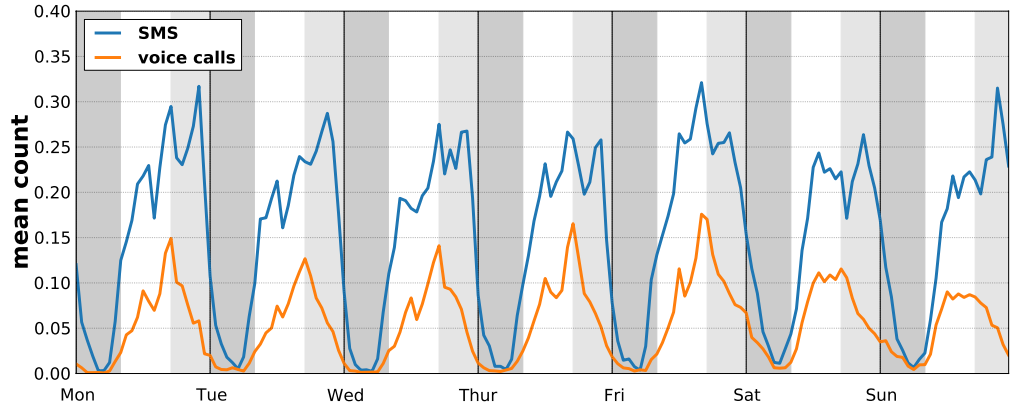}
\end{center}
\caption{
{\bf Weekly temporal dynamics of interactions.} All calls and SMS', both incoming and outgoing calculated over the entire dataset and averaged per user and per week, showing mean number of interactions users had in given weekly bin. Light gray denotes 5pm, the end of lectures at the university, dark gray covers night between 12am and 8am. SMS is used more for communication outside regular business hours.}
\label{fig:com_dynamics}
\end{figure}

Figure~\ref{fig:com_dynamics} shows the communication for SMS and voice calls (both incoming and outgoing, between participants and with the external world) as a time series, calculated through the entire year and scaled to denote the mean count of interactions users had in given hourly time-bins in the week.
Also here, we notice differences between those channels.
While both clearly show decrease in activity during lunch time, call activity peaks around the end of business day and drops until next morning; SMS on the other hand, after similar decrease, which we can associate with commute, displays another evening peak.
Also at night, SMS seems to be a more acceptable form of communication, with messages exchanges continuing late and starting early, especially on Friday night, when the party seems to never stop.

We point out that the call and SMS dynamics display patterns that are quite distinct from face-to-face interactions between participants showed in Figure~\ref{fig:dynamics}.
Although calls and SMS communication is different on the weekends, the difference is not as dramatic as in the face-to-face interactions between the participants.
This indicates that the face-to-face interactions we observe during the week are driven primarily by the university work, and only few of those ties manifest during the weekends, even as the participants are clearly socially active, sending and receiving calls and messages.

\begin{figure}[!ht]
\begin{center}
	\includegraphics[width=1\textwidth]{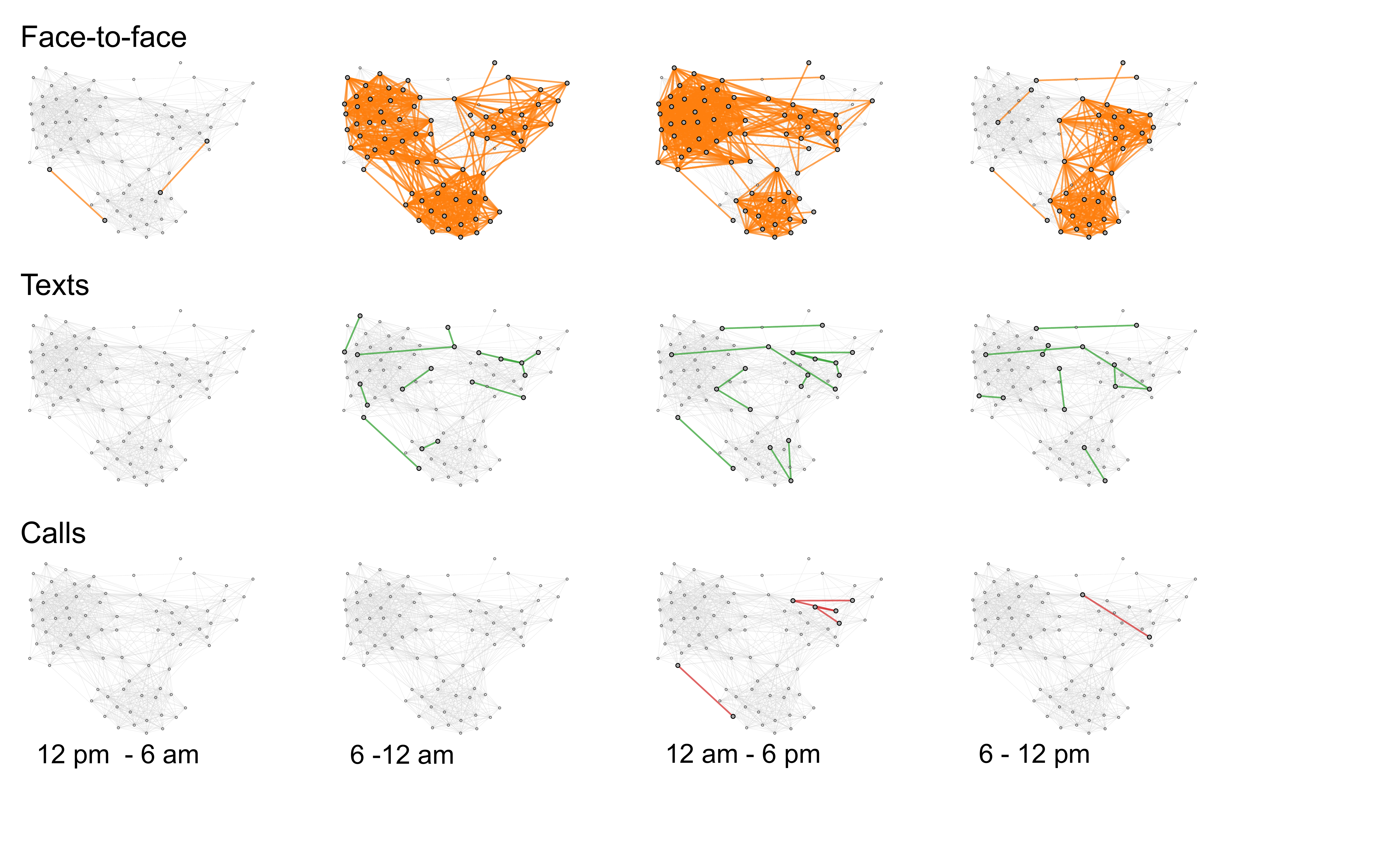}
\end{center}
\caption{
{\bf Daily activations in three networks.} One day (Friday) in a network showing how different views are produced by observing different channels.}
\label{fig:channel_dynamics}
\end{figure}

In Figure~\ref{fig:channel_dynamics} we focus on a single day (Friday) and show activation of links between participants in three channels: voice calls, text messages, and face-to-face meetings.
The three networks show very different views of the participants' social interactions.

\subsection{Online friendships}

\begin{figure}[!ht]
\begin{center}
	\includegraphics[width=1\textwidth]{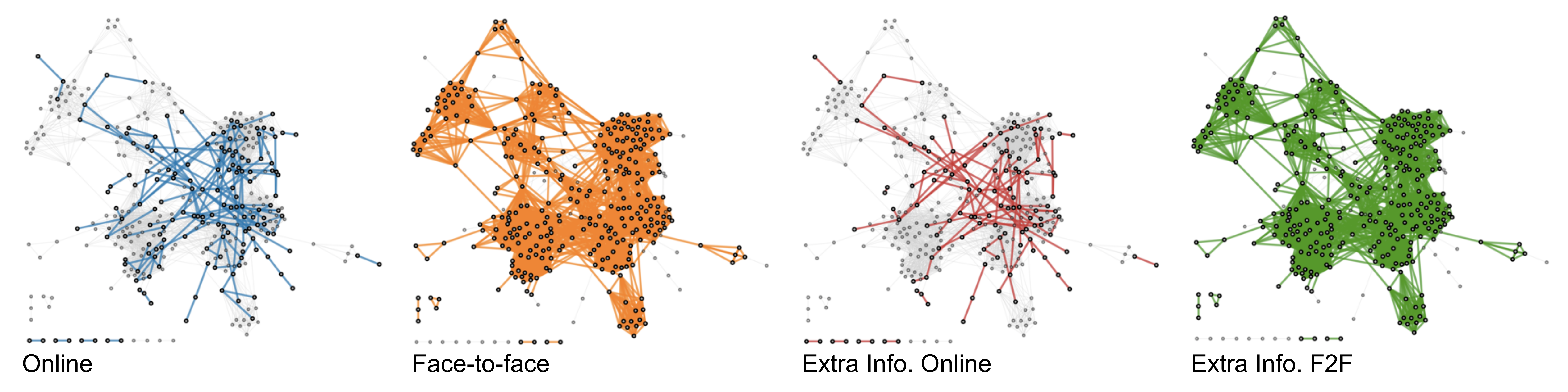}
\end{center}
\caption{
{\bf Face-to-face and online activity.} Figure shows data from the 2013 deployment for one representative week. 
\textbf{Online:} Interactions (messages, wall posts, photos, etc.) between users on Facebook. 
\textbf{Face-to-Face:} Only the most active edges, which account for $80\%$ of all traffic, are shown for clarity. 
\textbf{Extra Info. F2F:} Extra information contained in the Bluetooth data shown as the difference in the set of edges. 
\textbf{Extra Info. Online:} Additional information contained in the Facebook data.}
\label{fig:face_blue}
\end{figure}

The past years have witnessed a shift in our interaction patterns, as we have adapted new forms of online communication.
Facebook is to date the largest online social community with more than 1 billion users worldwide~\cite{facebookUsers}.
Collecting information about friendship ties and communication flows allows us to construct a comprehensive picture of the online persona.
Combined with other recorded communication channels we have an unparalleled opportunity to piece together an almost complete picture of all major human communication channels.
In the following section we consider Facebook data obtained from the 2013 deployment since, in contrast to the first deployment, we also collected interaction data.
For a representative week (Oct. 14 - Oct. 21, 2013) we collected 155 interactions (edges) between 157 nodes, yielding an average degree $\langle d \rangle=1.98$, average clustering $\langle c \rangle =0.069$, and average shortest path in the giant component (86 nodes) $\langle l \rangle = 6.52$.
The network is shown in the left most panel of Figure~\ref{fig:face_blue}.
By comparing with other channels we can begin to understand how well online social networks correspond to real life meetings.
The corresponding face-to-face network (orange) is shown in Figure~\ref{fig:face_blue}, where weak links, i.e. edges with fewer than 147 observations ($20\%)$ are discarded.
Corresponding statistics are for the 307 nodes and 3\,217 active edges: $\langle d \rangle=20.96$, $\langle c \rangle =0.71$, and $\langle l \rangle = 3.2$.
Irrespective of the large difference in edges, the online network still contains valuable information about social interactions which the face-to-face network misses---red edges in Figure~\ref{fig:face_blue}.

\subsection{Personality traits}

While the data from mobile sensing and online social networks provide insights primarily into structure of social ties, we are also interested in the demographics, psychological and health traits, and interests of the participants.
Knowing those characteristics, we can start answering questions about the reasons for the observed network formation; why are the ties created and what drives their dynamics?
For example, homophily plays a vital role in how we establish, maintain, and destroy social ties~\cite{mcpherson2001birds}.

Within the study, participants answered questions covering the aforementioned domains.
Those included a widely used \textit{Big Five Inventory}~\cite{john1999big} measuring five broad domains of human personality traits---openness, extraversion, neuroticism, agreeablenes, and conscientiousness.
The traits are scored on a 5-point Likert-type scale (low to high) and the average score of questions related to each personality domain are calculated. 

\begin{figure}[!ht]
\begin{center}
	\includegraphics[width=1\textwidth]{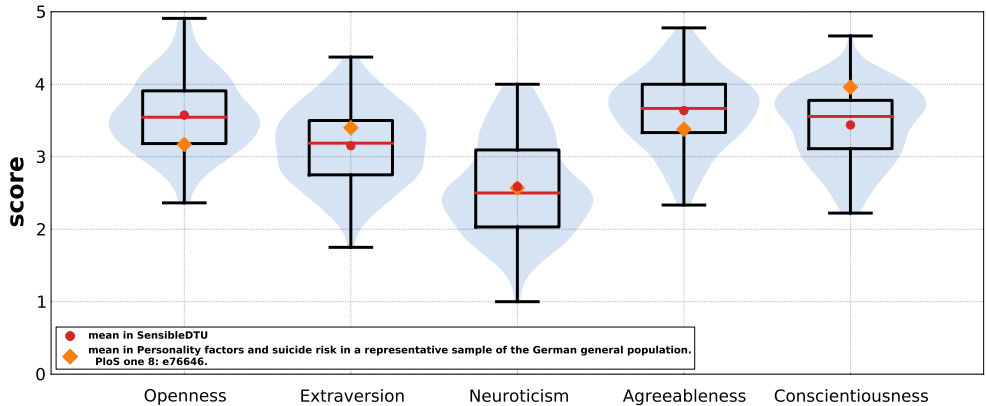}
\end{center}
\caption{
{\bf Personality traits} Violin plot of personality traits. Summary statistics are: {\bf openness} $\mu_O=3.58$, $\sigma_O=0.52$; {\bf extraversion} $\mu_E=3.15$, $\sigma_E=0.53$; {\bf neuroticism} $\mu_N=2.59$ $\sigma_N=0.65$; {\bf agreeablenes} $\mu_A=3.64$ $\sigma_A=0.51$; {\bf conscientiousness} $\mu_C=3.44$ $\sigma_C=0.51$. Mean values from our deployment (red circles) compared with mean values reported in~\cite{bluml2013personality} (orange diamonds).}
\label{fig:traits_summary}
\end{figure}

As Big Five has been collected for various populations, including representative sample from Germany~\cite{bluml2013personality} and students from western Europe~\cite{schmitt2007geographic}, we report the results from 2012 deployment in Figure~\ref{fig:traits_summary} to show that our population is unbiased with respect to those important traits.

\section{Perspectives}

We expect that the amount of data collected about human beings will continue to increase.
New and better services will be offered to the users, more effective advertising will be implemented, and researchers will be learning more about the human nature.
As the complexity and scale of studies on social systems studies grows, collection of high-resolution data for studying human behavior will become increasingly challenging on multiple levels, even when offset by the technical advancements.
Technical preparations, administrative tasks, and tracking data quality are a substantial effort for an entire team, before even considering the scientific work of data analysis.
It is thus an important challenge for the scientific community to create and embrace re-usable solutions, including best practices in privacy policies and deployment procedures, supporting technologies for data collection, handling, and analysis methods.

The results presented in this paper---while still preliminary considering the intended multi-year span of the project---clearly reveal that a single stream of data rarely supplies a comprehensive picture of human interactions, behavior, or mobility.
At the same time, creating larger studies, in terms of number of participants, duration, channels observed, or resolution, is becoming expensive using the current approach.
The interest of the participants depends on the value they get in return and the inconvenience the study imposes on their lives.
The inconvenience may be measured by decreased battery life of their phones, annoyance of answering questionnaires, and giving up some privacy.
The value, on the other hand, is classically created by material incentives, such as paying participants or, as in our case, providing smartphones and creating services for the participants.
Providing material incentives for thousands or millions of people, as well as related administrative effort of study management, may simply not be feasible.

In the not-so-distant future, many of the studies of human behavior will move towards accessing already existing personal data.
Even today we can access mobility of large populations, by mining data from Twitter, Facebook, or Flickr.
Or, with users' authorizations, we can track their activity levels, using API's of self-tracking services such as Fitbit or RunKeeper.
Linking across multiple streams is still difficult today (the problem of data silos), but as users take more control over their personal data, scientific studies can become consumers rather than producers of the existing personal data.

This process will pose new challenges and amplify the existing ones, such as replicability and reproducibility of the results or selection bias in the context of full end-user data control.
Still, we expect for the future studies to increasingly relay on the existing data, and it is important to understand how the incomplete view we get from such  data influences our results.
For this reason, we need research testbeds---such as the Copenhagen Networks Study---where we study `deep data' in the sense of multi layered data streams, sampled with high temporal resolution. These deep data will allow us to unlock and understand the future streams of big data.

\section{Acknowledgments}
The SensibleDTU project was made possible by a Young Investigator Grant from the Villum Foundation (\emph{High Resolution Networks}, awarded to SL). 
Scaling the project up to 1\,000 individuals in 2013 was made possible by a interdisciplinary UCPH 2016 grant, Social Fabric (PI David Dreyer Lassen, SL is co-PI) focusing mainly on the social and basic science elements of the project. 
This grant has funded purchase of the smartphones, as well as technical personnel. 
We are indebted to our University of Copenhagen partners on a number of levels:
All instrumentation on the 2013 questionnaires, as well as embedded anthropologist, are courtesy of the Social Fabric group, and most importantly the Social Fabric consortium has made valuable contributions through discussion and insight regarding nearly every aspect of the study.
For an overview of the Social Fabric project, see \texttt{http://socialfabric.ku.dk/}.
We thank group leaders in the Social Fabric Project: Professor David Dreyer Lassen, Professor Morten Axel Pedersen, Associate Professor Anders Blok, Assistant Professor Jesper Dammeyer, Associate Professor Joachim Mathiesen, Assistant Professor Julie Zahle, and Associate Professor Rikke Lund.
From Institute of Economics: David Dreyer Lassen, Andreas Bjerre Nielsen, Anne Folke Larsen, and Nikolaj Harmon.
From Institute of Psychology: Jesper Dammeyer, Lars Lundmann, Lasse Meinert Jensen, and Patrick Bender.
From Institute of Anthropology: Morten Axel Pedersen, My Madsen.
From Institute of Sociology: Anders Blok, Tobias Bornakke.
From Institute of Philosophy: Julie Zahle.
From Institute of Public Health: Rikke Lund, Ingelise Andersen, Naja Hulvej Rod, Ulla Christensen, and Agnete Skovlund Dissing.
From the Niels Bohr Institute (Physics): Joachim Mathiesen, and Mogens H\o{}gh Jensen.

\bibliography{bibliography,grant}



\end{document}